\documentclass[12pt]{article}
\usepackage{latexsym}
\usepackage{epsf}

\thicklines                         
\def\footnoteitem(#1)#2{
\begin{list}{#1}{\labelwidth4.0mm \leftmargin7.0mm
\labelsep2.5mm \rightmargin7.0mm \parsep0.5ex plus0.2ex minus0.1ex
\itemsep0ex plus0.2ex }
\item #2
\end{list}
}

\def\secteq#1{ \setcounter{equation}{0}
               \renewcommand{\theequation}{#1.\arabic{equation}} }

\begin{document}
\newcommand{\be}{\begin{equation}}
\newcommand{\ee}{\end{equation}}
\newcommand{\ba}{\begin{eqnarray}}
\newcommand{\ea}{\end{eqnarray}}

\newcommand{\cL}{{\cal L}}
\newcommand{\cM}{{\cal M}}
\newcommand{\Bt}{{\tilde B}}
\newcommand{\cO}{{\cal O}}
\newcommand{\cOt}{{\tilde\cO}}
\newcommand{\bt}{{\tilde\beta}}
\newcommand{\tr}{{\mbox{tr}\,}}
\newcommand{\str}{{\mbox{str}\,}}
\newcommand{\Exp}{{\mbox{exp}\,}}
\newcommand{\Mdot}{{\dot M}}
\newcommand{\Mbar}{{M_{VS}}}
\newcommand{\tb}{{\tilde\beta}}
\newcommand{\vp}{{\vec p}}
\newcommand{\hX}{{\hat X}}
\begin{titlepage}

\vskip 3mm
\rightline{UTCCP-P-88}
\rightline{June 2000}

\baselineskip=20pt plus 1pt
\vskip 0.5cm

\centerline{\LARGE On the Determination of Nonleptonic Kaon Decays}
\vskip 0.5cm
\centerline{\LARGE from $K\to\pi$ Matrix Elements}
\vskip 1.5cm
\centerline{\large Maarten Golterman$^1$}
\centerline{\em Center for Computational Physics, University
of Tsukuba, Tsukuba,}
\centerline{\em Ibaraki 305-8577, Japan}
\centerline{\em and}
\centerline{\em  Department of Physics, Washington University,
St. Louis, MO 63130, USA$^*$}
\vskip 0.5cm
\centerline{\large Elisabetta Pallante$^2$}
\centerline{\em Facultat de F\'{\i}sica, Universitat de Barcelona,
Diagonal 647, 08028 Barcelona, Spain}
\vskip 2.0cm
\baselineskip=12pt plus 1pt
\parindent 20pt
\centerline{\bf Abstract}
\textwidth=6.0truecm
\medskip
The coupling constants of the 
order $p^2$ low-energy weak effective
lagrangian can be determined from the $K\to\pi$
and $K\to 0$ weak matrix elements, choosing degenerate
quark masses for the first of these.  However, for
typical values of quark masses in Lattice QCD computations,
next-to-leading $O(p^4)$ corrections are too large to be ignored, and
will need to be included in future analyses.  Here we provide
the complete $O(p^4)$ expressions for these matrix elements
obtained from Chiral Perturbation Theory, valid for
partially quenched QCD with $N$ degenerate sea 
quarks.  Quenched QCD corresponds to the special case
$N=0$.  We also discuss the role of the $\eta'$ meson in
some detail, and we give numerical examples of the
size of chiral logarithms.

\frenchspacing

\nonfrenchspacing

\vskip 0.8cm

\noindent PACS: 13.25.Es, 12.38.Gc, 12.39.Fe

\vskip 0.8cm
\vfill
\noindent $^*$ permanent address \\
\noindent $^1$ e-mail: {\em maarten@aapje.wustl.edu}  \\
\noindent $^2$ e-mail: {\em pallante@ecm.ub.es}  \\
\end{titlepage}
\section{Introduction}
\secteq{1}
The determination of non-leptonic kaon-decay amplitudes from Lattice QCD
remains a challenging task.  However, recently there have been several
developments which may lead to substantial progress in this field,
ranging from new ideas on how to cope with chiral symmetry on the
lattice to a sizable increase of computer power available
for the necessary numerical computations.  

Still, there are important
theoretical difficulties afflicting the determination of the
relevant weak matrix elements, which are a consequence of the
fact that the final states contain more than one strongly
interacting particle (the pions).  This is formalized in what is
sometimes called the Maiani--Testa theorem \cite{mt}, which says
that it is not possible to extract the physical
matrix elements with the correct kinematics from the asymptotic
behavior in time of the euclidean correlation functions accessible to
numerical computation. 

There are several old and new ideas on the market on how to deal
with this situation, which divide into three groups.   First,
one may determine the matrix elements with a kaon in the
initial state and two (or more) pions in the final state for
an unphysical choice of external momenta.  From the large-time
behavior of the euclidean correlation function
\be
C(t_2,t_1)=\langle 0|\pi_1(t_2)\pi_2(t_2)\cO_{\rm weak}(t_1)K(0)|0\rangle
\;, \label{CORRF}
\ee
with the kaon at rest, one obtains the matrix element for
$\vp\equiv\vp_{\pi_1}=-\vp_{\pi_2}=0$ instead of the physical $|\vp|
=(1/2)\sqrt{m_K^2-4m_\pi^2}$, {\it i.e.} energy is not conserved
\cite{mt,gl}.  The idea is then to use chiral perturbation theory
(ChPT) in order to correct for this unphysical choice of momenta
\cite{bdhs}.  The most recent computation of the $\Delta I=3/2$
$K\to 2\pi$ matrix element using this method
can be found in Ref.~\cite{jlqcd}.  In this computation, all meson
masses were taken degenerate and the quenched approximation was
used.  Adjustments for all these
unphysical effects, which also include power-like
finite-volume effects coming from pion rescattering diagrams,
were made using one-loop ChPT \cite{gl}.
For recent ideas on choosing $m_K=2m_\pi$ for the
lattice computation (for which $\vp=0$ does conserve energy),
see Refs.~\cite{dawsonetal,gl2}.

For the $\Delta I=1/2$ case, the situation is more complicated
for a number of reasons. 
Here we only mention (since this is less well known) that
a quenched or partially quenched
computation appears to be afflicted by ``enhanced finite-volume
effects," which do not occur for the $\Delta I=3/2$ case.
This problem appears at one--loop in quenched 
ChPT \cite{mgeplat99,mgep}, but not much is known about this effect
beyond one loop.  We are investigating this issue.
For a nice review of many other issues, 
see Ref.~\cite{dawsonetal}.

A second idea was very recently proposed in Ref.~\cite{ll}, where it
was shown how the matrix elements of interest can in principle
be determined from finite-volume correlation functions without
analytic continuation.  The energy-conserving amplitude is
obtained by tuning the spatial volume such that the first
excited level of the two-pion final state has an energy 
equal to the kaon mass.  With sufficient accuracy
to determine the lowest excited levels, the finite-volume
matrix element may then be computed on the lattice, and subsequently be
converted into the physical infinite-volume
amplitude.  For this, it is obviously necessary to choose meson masses
such that $2m_\pi<m_K$ (as well as $m_K<4m_\pi$,
so that the final-state pions are in the elastic regime). 
Again, if such a computation is done in a (partially) quenched setting
one might expect that enhanced finite-volume effects 
could also occur with this method.

A third idea is based on the observation that, if one needs
ChPT anyway in order to convert an unphysical matrix element
into a physical one, one might as well choose the unphysical
matrix element as simple as possible.  Chiral symmetry relates
the $K\to 2\pi$ matrix elements of interest to the simpler $K\to\pi$
and $K\to$~vacuum ($K\to 0$) 
matrix elements of the same weak operators
which mediate non-leptonic kaon decay \cite{bernardetal}.
Advantages of this approach are that there are no strongly
interacting particles in the final state, and that lattice
computations of these simpler matrix elements may be less
difficult.  

The first advantage is, in a sense, not really an advantage if one 
wishes to convert the results of a lattice computation into
a calculation of the non-leptonic kaon decay rates, because
final-state interactions will still have to be taken into
account.   However, this method does avoid all the {\it
unphysical} effects, such as 
power-like or even enhanced finite-volume effects, associated
with the multi-pion final state.  Formulated in another way, 
with this method the simplest possible matrix elements
(in this case $K\to\pi$ and $K\to 0$) are used to obtain the
relevant weak low-energy constants (LECs) of the weak effective
lagrangian.  Using ChPT, these can then be converted into
estimates of the kaon-decay rates. 

A key question is which order in ChPT will be
needed in order to carry out such a program.   At tree level
({\it i.e.} $O(p^2)$), only three LECs come into play, but at one loop
({\it i.e.} $O(p^4)$) many more LECs contribute to all relevant matrix
elements \cite{kmw}.   In fact, from the analysis reported
in this paper as well as from previous work it is 
clear that tree-level ChPT is not enough \cite{gl,mgeplat99,pallante}.
In addition (as we will demonstrate), not all $O(p^4)$
LECs needed for $K\to 2\pi$ decays can be obtained from
$K\to\pi$ and $K\to 0$ matrix elements.
However, as we will advocate in this paper, it may be
possible to determine at least the $O(p^2)$ LECs from
a lattice computation, taking one-loop ChPT effects  
into account.   A  reliable, first-principle determination
of the $O(p^2)$ octet and 27-plet LECs would clearly be
interesting by itself.  Moreover, phenomenological estimates
of these LECs, based on a one-loop ChPT analysis of experimental
data are available \cite{kmw1}, making a direct comparison 
possible.

In this paper we present an analysis of $K\to\pi$ and $K\to 0$
amplitudes in one-loop ChPT, with the above described
philosophy in mind.  For $K\to\pi$ we choose our valence quark masses
to be degenerate, thus conserving energy for this case.
The analysis is performed in partially
quenched ChPT \cite{bgpq}, with an arbitrary number of degenerate
sea quarks.  We present the results for these matrix elements
in terms of the quark masses.  

In Sect.~2, we list and discuss all $O(p^2)$ and
$O(p^4)$ operators needed for our calculation, including those
containing the $\eta'$ meson.  In Sect.~3, we discuss
the role of the $\eta'$ in partially quenched QCD in some more
detail than has been done so far in the literature. This
section can be skipped if one is only interested in results.
In Sect.~4, we give complete one-loop expressions for the
octet and 27-plet $K\to\pi$ and $K\to 0$ matrix elements,
including contributions from $O(p^4)$ operators, organized
by subsection.  In Subsect.~4.1 partially quenched results
for $N$ degenerate sea quarks are presented, which are valid
also in the case that the meson made out of sea quarks 
(the ``sea meson") is not
light compared to the $\eta'$ (a realistic
situation in actual lattice computations).  In Subsect.~4.2
we specialize to the case that the sea meson is light compared
to the $\eta'$.  Subsect.~4.3 contains the completely quenched
results, obtained by setting $N=0$ and keeping the $\eta'$.
For completeness, we include the fully unquenched
results, with non-degenerate sea-quark masses for the $K\to 0$
matrix elements, in Subsect.~4.4.  
In Sect.~5 we present a detailed discussion
of the results, including the role of $O(p^4)$ operators and 
 numerical examples for typical choices of the
parameters.  The last section contains our conclusions.

\bigskip
\section{Definition of operators}
\secteq{2}
Partially quenched QCD may be defined by separately introducing
valence- and sea-quark fields, each with their own mass.  The
valence quarks are quenched by introducing for each valence
quark a ``ghost" quark, which has the same mass and quantum numbers,
but opposite statistics \cite{bgpq,morel}.  This, in effect, 
removes the valence-quark determinant from the QCD partition
function.  We will consider a theory with $n$ quarks, of which $N$
are sea quarks, and $n-N\ge 3$ valence quarks.  This requires $n-N$
ghost quarks, with masses equal to those of the valence quarks.  
We will consider valence quarks with arbitrary masses
$m_1,\dots,m_{n-N}$, and degenerate sea quarks, all with mass $m_S$.
The relevant chiral symmetry group is the graded group
SU($n|n-N$)$_L\otimes$SU($n|n-N$)$_R$ \cite{bgpq}.  Fully quenched
QCD arises as a special case of this construction by taking $N=0$ 
\cite{bgq}, or equivalently, when the sea quarks are decoupled by 
taking $m_S\to\infty$.

The euclidean low-energy effective lagrangian  which mediates non-leptonic 
weak transitions with $\Delta S=1$ is given by
\be
\cL_{\Delta S=1}=\cL_2+\cL_2^{\eta'}+\cL_4+\ldots\;,\label{LWEAK}
\ee
where the dots denote higher order terms in the chiral expansion.
The  $O(p^2)$ lagrangian \cite{bernardetal} $\cL_2$ contains three terms
\footnote{In the analysis of CP conserving weak amplitudes one can safely
disregard $O(e^2p^0)$ terms induced by electroweak interactions.}
\be
\cL_2=-\alpha^8_1\,\str(\Lambda L_\mu L_\mu)+\alpha^8_2\,\str(\Lambda X_+)
+\alpha^{27}\,T^{ij}_{kl}(L_\mu)^k_{\ i}(L_\mu)^l_{\ j}+{\mbox{h.c.}}
\;,\label{LTWO}
\ee
where $\str$ denotes the supertrace in flavor space.
Note that the supertraces become normal traces, $\str\to\tr$, in the case
$N=n$.
The terms with couplings $\alpha^8_1,\alpha^8_2$ transform as $(8_L,1_R)$, 
while the term with coupling $\alpha^{27}$ transforms as $(27_L,1_R)$.
The order $p^2$ lagrangian $\cL_2^{\eta'}$ will be discussed toward the
end of this section. The  $O(p^4)$ lagrangian can be written as
\be
\cL_4=\frac{1}{(4\pi f)^2}\left(
\sum_i\beta^8_i\,\cO^8_i+\sum_i\bt^8_i\,\cOt^8_i
+\sum_i\beta^{27}_i\,\cO^{27}_i\right)\;,\label{LFOUR}
\ee
with $(8_L,1_R)$ $O(p^4)$ operators $\cO^8_i$, $\cOt^8_i$
and  $(27_L,1_R)$  $O(p^4)$ operators
$\cO^{27}_i$ \cite{kmw,pallante,ekw}.  
The operators $\cOt^8_i$ denote total-derivative
operators; they do not contribute to energy-momentum conserving
matrix elements. However, they do contribute to the $K\to 0$ matrix element,
which does not conserve energy for non-degenerate quark masses.  Note that
there are no 27-plet total-derivative operators that contribute to the 
matrix elements considered in this paper, to $O(p^4)$.
The fields entering the weak operators are defined as follows
\ba
L_\mu&=&i\Sigma\partial_\mu\Sigma^\dagger\;, \label{OPS} \\
X_\pm&=&2B_0(\Sigma M^\dagger\pm M\Sigma^\dagger)\;, \nonumber 
\ea
where $B_0$ is the parameter $B_0$ of Ref.~\cite{gasleu}
and $B_0=4v/f^2$ in the notation of  Ref.~\cite{bernardetal}.
The unitary field $\Sigma$ is defined in terms of the hermitian
field $\Phi$ describing the Goldstone meson multiplet as
\be
\Sigma=\Exp(2i\Phi/f)\;, \label{SIGMA}
\ee
where $f$ is the bare pion-decay constant, normalized such that
$f_\pi=132$~MeV.
Note that $L_\mu$ and $X_\pm$ transform as $(8_L,1_R)$ under
(valence-flavor) SU(3)$_L\otimes$SU(3)$_R$.
The matrix $\Lambda$ in the lagrangian (\ref{LTWO}) 
picks out the $\Delta S=1$, 
$\Delta D=-1$ part of the octet operators, all with 
$\Delta I=1/2$:
\be
\Lambda^i_{\ j}=\delta^{i3}\delta_{j2}\;, \label{LAMBDA}
\ee 
where $i,j=1,2,3\ldots $ (or $u,d,s\ldots$) denote valence flavors.
The tensor $T^{ij}_{kl}$ projects onto the $\Delta I=1/2$ 
part of the $(27_L,1_R)$ operator, with nonzero components
\ba
T^{13}_{12}&=&T^{31}_{12}=T^{13}_{21}=T^{31}_{21}=\frac{1}{2}\;,
\label{HALF} \\
T^{23}_{22}&=&T^{32}_{22}=1\;, \nonumber \\
T^{33}_{32}&=&T^{33}_{23}=-\frac{3}{2}\;, \nonumber
\ea
or onto the $\Delta I=3/2$ part, with nonzero components
\ba
T^{13}_{12}&=&T^{31}_{12}=T^{13}_{21}=T^{31}_{21}=\frac{1}{2}\;,
\label{THREEHALF} \\
T^{23}_{22}&=&T^{32}_{22}=-\frac{1}{2}\;. \nonumber
\ea
The term with coupling $\alpha^8_2$ is known as the ``weak mass term,"
and mediates the $K\to 0$ transition at tree level. Its odd-parity
part, which in principle can also
contribute to the octet $K\to \pi\pi$ amplitude,
is proportional to $m_s -m_d$. For $m_s\neq m_d$ the weak mass term is a
total derivative \cite{bernardetal,bpp}, 
and therefore does not contribute to any energy-momentum-conserving
physical matrix element, like $K\to \pi\pi$.  
Instead, the $K\to 0$ and, for $M_K\ne M_\pi$,
$K\to\pi$ matrix elements do not conserve energy, and therefore the weak mass
term does contribute to both of them.
For $m_s=m_d$ the weak mass term  is not a total derivative, so that
it contributes also to the $K\to\pi$ matrix element with $M_K=M_\pi$.
Hence, in order to determine the octet coupling $\alpha^8_1$ from a computation
of the $K\to\pi$ matrix element, another quantity such as the $K\to 0$
matrix element is always needed, in order to eliminate the dependence on
 $\alpha^8_2$.

At order $p^4$ there are eight $(8_L,1_R)$ operators and six $(27_L,1_R)$
operators which can contribute  
to $K\to 0$, $K\to\pi$ and $K\to\pi\pi$ matrix elements.  The octet operators
can be written as follows 
\ba
\cO^8_1&=&\str(\Lambda X_+X_+)\;,\label{OCTET} \\
\cO^8_2&=&\str(\Lambda X_+)\,\str(X_+)\;,\nonumber \\
\cO^8_3&=&\str(\Lambda X_-X_-)\;,\nonumber \\
\cO^8_5&=&\str(\Lambda[X_+,X_-])\;,\nonumber \\
\cO^8_{10}&=&\str(\Lambda\{X_+,L_\mu L_\mu\})\;,\nonumber \\
\cO^8_{11}&=&\str(\Lambda L_\mu X_+ L_\mu)\;,\nonumber \\
\cO^8_{13}&=&\str(\Lambda X_+)\,\str(L_\mu L_\mu)\;,\nonumber \\
\cO^8_{15}&=&\str(\Lambda[X_-,L_\mu L_\mu])\;,\nonumber 
\ea 
while the 27-plet operators are
\ba
\cO^{27}_1&=&T^{ij}_{kl}(X_+)^k_{\ i}(X_+)^l_{\ j}\;,\label{TSPLET} \\
\cO^{27}_2&=&T^{ij}_{kl}(X_-)^k_{\ i}(X_-)^l_{\ j}\;, \nonumber \\
\cO^{27}_4&=&T^{ij}_{kl}(L_\mu)^k_{\ i}(\{L_\mu,X_+\})^l_{\ j}\;, \nonumber \\
\cO^{27}_5&=&T^{ij}_{kl}(L_\mu)^k_{\ i}([L_\mu,X_-])^l_{\ j}\;, \nonumber \\
\cO^{27}_6&=&T^{ij}_{kl}(X_+)^k_{\ i}(L_\mu L_\mu)^l_{\ j}\;, \nonumber \\
\cO^{27}_7&=&T^{ij}_{kl}(L_\mu)^k_{\ i}(L_\mu)^l_{\ j}
\,\str(X_+)\;. \nonumber 
\ea
These operators are the same as those in Ref.~\cite{bpp},
apart from the replacement $\tr\to\str$.  
For the energy-momentum non-conserving matrix elements the only 
total-derivative term that is needed is
\be
\cOt^8_1=i\partial_\mu\,\str(\Lambda[L_\mu , X_+ ])\;.
\label{OCTETTOTDER}
\ee

In general,
in the (partially) quenched formulation of the effective theory 
one needs to keep the $\eta'$ \cite{bgq,glpq,sharpecapri,sharpeq}, 
defined 
as the SU($n|n-N$)$_L\otimes$SU($n|n-N$)$_R$ invariant \cite{bgpq}
\be
\eta'=\str(\Phi)\;. \label{ETAP} 
\ee  
Note that this normalization  differs
from the one in Ref.~\cite{bgpq}, but is more convenient in keeping
track of  $N$ dependence. The presence of the $\eta '$
leads to new operators in the strong \cite{bgq,gasleu} and weak effective 
lagrangians.  Under CPS symmetry \cite{bernardetal}
all operators in $\cL_{2,4}$ of Eq. (\ref{LWEAK}) 
are even, and, since $\eta'$ is CPS odd,
new weak operators can be constructed by multiplying $\cL_{2,4}$ by even 
powers of $\eta'$.  It turns out that
such operators do not contribute to the quantities of interest 
in this paper.  However, other new weak operators arise from
multiplying CPS-odd weak operators   
by odd powers of $\eta'$. There are two operators of interest at order $p^2$,
so that $\cL_2^{\eta'}$ is given by
\be
\cL_2^{\eta'}=\gamma^8_1\,\partial_\mu(\eta'/f)\str(\Lambda L_\mu)
+i\gamma^8_2\,(\eta'/f)\str(\Lambda X_-)\;. \label{SPECIAL}
\ee
Since we are not interested in processes with external $\eta'$ lines, we 
do not consider new operators of order $p^4$ containing the $\eta'$ field.
 
For $N=3$ sea quarks the dynamics of the partially quenched theory is precisely
that of unquenched QCD with degenerate quark masses \cite{bgpq}.
Since all the low-energy constants (LECs) $\alpha^{8}_i,\, \alpha^{27}$,
$\gamma^8_i$ and $\beta^{8,27}_i,\, \tilde\beta_i^8$ (together with the 
 strong counterterms)
are independent of quark masses,
it follows that their $N=3$ partially-quenched values are equal to
those of the real world \cite{sharpepq}.  
However,  for
this equivalence to be valid, the $\eta'$ should be treated in the same way 
in the partially quenched theory as in the real world, so that 
one has to consider the limit in which the $\eta'$ decouples.
\section{The role of the $\eta'$ in partially quenched ChPT}
\secteq{3}
Before we present our results, we would like to discuss the
role of the $\eta'$ in a partially quenched theory
in more detail.  First, define the bare
(or tree-level) meson masses
\be
M^2_{ij}=B_0(m_i+m_j)\;,~~~~~~~i,j=1,\ldots 2n-N\;, \label{MMASS}
\ee
for a light pseudoscalar (pseudo-Goldstone) meson made out of quarks 
or ghost quarks $i$ and $j$.  For degenerate
sea quarks, this simplifies to $M^2_{SS}=2B_0m_S$.  The 
two-point function for neutral mesons $\Phi_{ii}$ (in that basis)
 is given by \cite{bgpq}
\ba
\langle\Phi_{ii}(x)\Phi_{jj}(0)\rangle&=&
\int\frac{d^4p}{(2\pi)^4}\,e^{ipx}\,G_{i,j}(p)\;, \nonumber \\
G_{i,j}(p)&=&\frac{\delta_{ij}\epsilon_i}{p^2+M^2_{ii}}
-X_{ij}(p)\;, \label{GIJ} \\
X_{ij}(p)&=&\frac{1}{3+N\alpha}\;
\frac{(m_0^2+\alpha p^2)(p^2+M^2_{SS})}
{(p^2+M^2_{ii})(p^2+M^2_{jj})(p^2+M^2_{\eta'})}\;,
\nonumber
\ea
where
$$
\epsilon_i =\cases{ +1,&for $1 \le i \le n$ \cr
-1,&for $n+1 \le i \le 2n-N$ \cr}
$$
and the $\eta'$ mass is given by
\be
M^2_{\eta'}=\frac{M^2_{SS}+Nm_0^2/3}{1+N\alpha/3}\;.
\label{ETAPMASS}
\ee
The parameters $m_0^2$ and $\alpha$ (not to be confused with
$\alpha^{8,27}_i$) come from the strong-lagrangian $O(p^2)$
operators quadratic in the $\eta'$ field, $(Nm_0^2/6)(\eta')^2
+(N\alpha/6)(\partial_\mu\eta')^2$.  
The term $X_{ij}(p)$ has a double pole for  $M_{ii}=M_{jj}$, unless
$M_{SS}=M_{ii}=M_{jj}$.  This implies that 
partially quenched theories suffer from the same ``quenched
infrared diseases" as the quenched theory unless all valence-quark
masses are equal to the sea-quark mass \cite{bgpq}.  
{}From Eq.~(\ref{GIJ}) it is
easily verified that the $\eta'$ two-point function is just
$(N/(1+N\alpha/3))(p^2+M^2_{\eta'})^{-1}$.

The quantity $X_{ij}(p)$ can also be written as
\ba
X_{ij}(p)&=&\frac{1}{3}\left(\frac{A_{ij}}{p^2+M^2_{ii}}
\,-\,\frac{B_{ij}}{p^2+M_{\eta'}^2}
+\frac{\cM^2_{ij}-A_{ij}M^2_{jj}}{(p^2+M^2_{ii})(p^2+M^2_{jj})}
\right)\;, \label{X}
\ea
with 
\ba
\cM^2_{ij}&=
&\frac{(N/3)(m_0^2-\alpha M_{ii}^2)(m_0^2-\alpha M_{jj}^2)M_{SS}^2
+m_0^2(M_{SS}^2-M_{ii}^2)(M_{SS}^2-M_{jj}^2)}
{[(N/3)(m_0^2-\alpha M_{ii}^2)+M_{SS}^2-M_{ii}^2]
[(N/3)(m_0^2-\alpha M_{jj}^2)+M_{SS}^2-M_{jj}^2]}\;, \nonumber \\
A_{ij}&=&\frac{(N/3)(m_0^2-\alpha M_{ii}^2)(m_0^2-\alpha M_{jj}^2)
+\alpha(M_{SS}^2-M_{ii}^2)(M_{SS}^2-M_{jj}^2)}
{[(N/3)(m_0^2-\alpha M_{ii}^2)+M_{SS}^2-M_{ii}^2]
[(N/3)(m_0^2-\alpha M_{jj}^2)+M_{SS}^2-M_{jj}^2]} \;, \nonumber \\
B_{ij}&=&
\frac{(N/3)(m_0^2-\alpha M_{SS}^2)^2/(1+\alpha N/3)}
{[(N/3)(m_0^2-\alpha M_{ii}^2)+M_{SS}^2-M_{ii}^2]
[(N/3)(m_0^2-\alpha M_{jj}^2)+M_{SS}^2-M_{jj}^2]} \;.
\nonumber \\
\label{MAB}
\ea
The coefficients $A_{ij}$, $B_{ij}$ and $\cM^2_{ij}$ are complicated 
functions of
the various mass scales in the partially quenched effective theory.  
We may consider various limits in which these expressions simplify
considerably.  First, one easily
obtains the fully quenched expression by setting $N=0$, or equivalently 
taking $M_{SS}\to\infty$, finding for all $ij$
\be
\cM^2_{ij}\to m_0^2\;,\ \ \ A_{ij}\to\alpha\;,\ \ \ B_{ij}
\to 0\;. \label{CALMQLIMIT}
\ee
It is clear from these expressions that in the quenched case, the
$\eta'$ should be kept in the effective theory.
Another interesting limit is
that in which the $\eta'$ decouples \cite{sharpepq} (which, as inspection of
Eq.~(\ref{ETAPMASS}) tells us, is only possible for $N>0$).
In this limit, again for all $ij$,
\be
\cM^2_{ij}\to\frac{3}{N}M_{SS}^2\;,\ \ \ A_{ij}
\to\frac{3}{N}\;, \label{CALMLIMIT}
\ee
and we drop the $\eta'$ pole in Eq.~(\ref{X}). 
The dependence of $X_{ij}$ on the $\eta'$ parameters has 
disappeared in this limit.

As argued in Ref.~\cite{glpq}, in actual 
partially quenched Lattice
QCD computations, the sea-meson mass $M_{SS}$ maybe comparable in
size to $m_0$ so that the full dependence
of $\cM^2_{ij}$ and $A_{ij}$ on the parameters $m_0$, $\alpha$ and $M_{SS}$
should be kept.  A third possibility is then
given by the limit in which the valence-meson mass is 
small compared to the $\eta'$ mass, {\it i.e.} $M_{kk}\ll M_{\eta'}$.
The expressions for $\cM^2_{ij}$, $A_{ij}$ and $B_{ij}$ given in Eq.
(\ref{MAB}) reduce to
those of Ref.~\cite{glpq} if we expand in $M_{kk}^2/M_{\eta'}^2$
but not in $M_{SS}^2/M_{\eta'}^2$:
\ba
\cM^2_{ij}&\to&\cM^2=\frac{m_0^2M_{SS}^2}{(N/3)m_0^2+M_{SS}^2}
\left(1+O\left(\frac{M_{kk}^4}{M_{\eta'}^4}\right)\right)\;, 
\label{XLIMIT} \\
A_{ij}&\to&A=\frac{(N/3)m_0^4+\alpha M_{SS}^4}
{[(N/3)m_0^2+M_{SS}^2]^2}+O\left(\frac{M_{kk}^2}{M_{\eta'}^2}\right)
\;, \nonumber \\
B_{ij}&\to&\frac{(N/3)(m_0^2-\alpha M_{SS}^2)^2/(1+\alpha N/3)}
{[(N/3)m_0^2+M_{SS}^2]^2}
+O\left(\frac{M_{kk}^2}{M_{\eta'}^2}\right)\;. \nonumber
\ea
The (partially) quenched expansion we consider in this paper is
systematic if we take $\cM^2$ to be of order $p^2$, in other words,
if we take the parameter $m_0^2$ to be of the same order as the 
quark mass, just as in the case of quenched ChPT~\cite{bgq,glpq,cp}.
  
It was also
shown in Ref. \cite{glpq} that, for the quantities considered 
there, simply
ignoring one-loop contributions coming from the $\eta'$ pole
in Eq.~(\ref{X}) and then taking the limit $M_{\eta'}\to\infty$ in the rest
is the same as matching to the limit in which
the $\eta'$ decouples.  In other words, if we ignore these
contributions, the LECs appearing in those quantities are the
same in the $N=3$ partially quenched world and the real world.
In addition, when we take only $M_{\eta'}/M_{kk}$ large, but not
$M_{\eta'}/M_{SS}$, these one-loop contributions are polynomial
in the valence-meson masses, and can still be ignored if we are
interested in the non-analytic dependence on the 
valence-meson masses.

The same observations are also true here,
even though there are diagrams with a more complicated topology
than the simple tadpole diagrams needed in Ref.~\cite{glpq}.
For $K\to\pi$,
there are contributions of the form depicted in Fig. 1.
\begin{figure}
\begin{center}
\leavevmode\epsfxsize=3.5cm\epsfysize=1.5cm\epsfbox{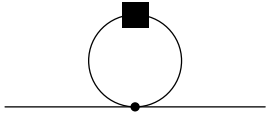}
\end{center}
\caption{A diagram contributing to the $\Delta S=1$ $K\to\pi$
matrix element.
A line is a pseudoscalar meson propagator, a dot a strong vertex and a box is 
a weak vertex.
\label{figure1}}
\end{figure}
Taking $M=M_{kk}$ to be the degenerate valence-meson mass running in the loop
and abbreviating $B=B_{kk}$, this diagram leads to one-loop integrals such as
\be
\int\frac{d^4p}{(2\pi)^4}\;\frac{1}{p^2+M^2}\;
\frac{B}{p^2+M^2_{\eta'}}
=\frac{B}{M_{\eta'}^2-M^2}
\int\frac{d^4p}{(2\pi)^4}\left(\frac{1}{p^2+M^2}
-\frac{1}{p^2+M^2_{\eta'}}\right)\;, \label{INT}
\ee
multiplied by two powers of $M^2$ from the two
$O(p^2)$ vertices in the diagram (there are no contributions
from vertices proportional to $M_{SS}^2$, so the only
dependence on $M_{SS}$ comes through the coefficient $B$).  
The integral contributes
an $\eta'$ chiral logarithm, which we drop as above, and a
Goldstone chiral logarithm proportional to $BM^6/(M_{\eta'}^2-M^2)\cdot
\log{M^2}$.  If we do not assume that
$M$ is small compared to $M_{\eta'}$ this is of order $M^4$,
but if we expand in $M^2/M_{\eta'}^2$,
this constitutes an $O(p^6)$ contribution.

In all the following calculations, we will assume that valence-meson masses are
sufficiently small compared to $M_{\eta'}$ to justify the
expansion in $M^2/M_{\eta'}^2$.  If one would not make this
assumption, all contributions from integrals like Eq.~(\ref{INT})
would have to be kept, since both the Goldstone-meson and the
$\eta'$ pole give rise to additional non-polynomial $M$ dependence.
However, with this assumption, contributions coming from $\eta'$
tadpoles or diagrams such as Fig.~1 containing an $\eta'$ on the loop
are analytic in the valence-meson masses to order $p^4$, and we will therefore
drop them from consideration.

Finally, we note that the coefficients of chiral logarithms will
in general
depend in a complicated non-polynomial way on the valence- and
sea-meson masses, so that the  $O(p^4)$ LECs cannot be defined in a
mass-independent way.  The coefficients of  chiral logarithms have a
polynomial dependence on meson masses
only if we expand in {\it both} $M^2/M_{\eta'}^2$ and
$M_{SS}^2/M_{\eta'}^2$.  In that case
the LECs of the partially quenched theory without the $\eta'$
(and with $N=3$) are the same as those of the real world.

\section{$K\to\pi$ and $K\to 0$ matrix elements at $O(p^4)$}
\secteq{4}
In this section we calculate $K^0\to 0$ and $K^+\to\pi^+$ matrix elements
at order $p^4$ in the effective theory. Four cases are considered: partially
quenched with the valence-meson mass $M\ll M_{\eta'}$ but
$M_{SS}/M_{\eta'}$ arbitrary, partially quenched with also
$M_{\eta'}/M_{SS}$ large, quenched, and unquenched.
\subsection{Partially quenched results}

We consider a partially quenched theory with three valence quarks, $u$, $d$
and $s$ and $N$ degenerate sea quarks. The matrix elements to be calculated 
are defined as
\ba
[K^0\to 0]
&\equiv&\langle 0|\cL_{\Delta S=1}|K^0\rangle\;, \label{KVAC} \\ 
{}[K^+\to\pi^+]
&\equiv&\langle\pi^+|\cL_{\Delta S=1}|K^+\rangle\;. \nonumber
\ea
The first process is calculated for $m_s\neq m_d$ and $m_u=m_d$.
In the last process we take  the valence-quark masses all equal $m_s=m_d=m_u$,
so that it conserves energy.  In this SU(3) limit, the $K^0\to\eta$ matrix
element does not contain any extra information.
 
We have calculated these matrix elements to $O(p^4)$ in partially quenched 
ChPT, using
dimensional regularization in the $\overline{MS}$ scheme, and
including contributions from the $O(p^4)$ operators 
(\ref{OCTET}) and (\ref{TSPLET}).  In the case with degenerate
valence quarks,
let $M=M_{ii}$ be the physical mass of a meson made out of valence quarks, 
$M_{SS}$ that of a meson made out of sea quarks, and $\Mbar$ that
of a meson made out of a valence and a sea quark.  At tree level
in ChPT one has 
\be
\Mbar^2=\frac{1}{2}(M^2+M_{SS}^2)\;.
\ee
Define, for any mass M,
\be
L(M)=\log\frac{M^2}{\Lambda^2}\;,\label{LOG}
\ee
where $\Lambda$ is the $\overline{MS}$ scale.  In addition, define,
for any two masses $M_1$ and $M_2$,
\be
L_n(M_1,M_2)=\frac{M_1^{2n}\log\frac{M_1^2}{\Lambda^2}
-M_2^{2n}\log\frac{M_2^2}{\Lambda^2}}{M_1^2-M_2^2}\;.\label{LNS}
\ee

As discussed in the previous section, we will
assume that the valence-meson masses are small compared to
$M_{\eta'}$ ({\it cf.} Eq.~(\ref{ETAPMASS})), but not make the same
assumption about $M_{SS}$.
At one loop, we then have, for the octet $K\to\pi$ matrix element, 
using $\cM^2$ and $A$ from Eq.~(\ref{XLIMIT}),
\ba
&&\hspace{-0.7truecm}[K^+\to\pi^+]_8=\frac{4\alpha^8_1M^2}{f^2}\left(1-
\frac{1}{(4\pi f)^2}\Biggl[N\Mbar^2(L(\Mbar)-1)
\right.
\label{KPIOCTET} \\
&&\hspace{2.4truecm} \left.
+\,2\left(\cM^2-\frac{8}{3}AM^2\right)L(M)
+\frac{2}{3}(\cM^2+2AM^2)\Biggr]\right) \nonumber \\
&&\hspace{2.0truecm}
-\,\frac{4\alpha^8_2M^2}{f^2}\left(1-\frac{1}{(4\pi f)^2}\left[
\frac{2}{3}(\cM^2-4AM^2)L(M)+\frac{2}{3}\cM^2
\right]\right)
\nonumber \\
&&\hspace{1.0truecm}
+\frac{4(\gamma^8_1+2\gamma^8_2)M^2}{f^2(4\pi f)^2}\;
\Biggl[M^2(2L(M)-1)
-\frac{1}{3}NA\left(L_2(M_{SS},M)-M_{SS}^2-M^2\right) \nonumber \\
&&\hspace{4.3truecm}
-\frac{N}{3}\,\frac{\cM^2-AM^2}{M_{SS}^2-M^2}
\left(2M^2L(M)-L_2(M_{SS},M)+M_{SS}^2\right)
\Biggr]\;,\nonumber 
\ea 
and for the 27-plet $K\to\pi$ matrix element,
\ba
&&\hspace{-0.7truecm}[K^+\to\pi^+]_{27}=
-\frac{4\alpha^{27}M^2}{f^2}\left(\!1-\frac{1}{(4\pi f)^2}
\Biggl[2N\Mbar^2(L(\Mbar)-1)+2M^2(3L(M)-2)\right.\nonumber\\
&&\hspace{6truecm}\left.
+\frac{2}{3}(\cM^2-2AM^2)L(M)+\frac{2}{3}AM^2
\Biggr]\!\right) .
\label{KPITSPLET}
\ea
In the latter case, the matrix elements for $\Delta I=1/2$
and $\Delta I=3/2$ are the same,
because of SU(3) symmetry.  The $\eta'$ does not contribute
directly to the 27-plet matrix elements; the dependence on
$m_0^2$ and $\alpha$ comes from the fact that we expressed all
results in terms of bare meson masses, or, equivalently,
in terms of quark masses ({\it cf.} Eq.~(\ref{MMASS})).  
The octet matrix element receives instead direct contributions from $\eta'$
exchange.
Note that the
contributions from the two $O(p^2)$ $\eta'$-operators of Eq.~(\ref{SPECIAL})
have the same form.  
This is explained by the fact that, after a partial
integration, the first term in Eq.~(\ref{SPECIAL}), using the
equation of motion for $\Sigma$, is proportional to the
second term.
 
For the contributions from the $O(p^4)$ operators, we find
\ba
(4\pi f)^2[K^+\to\pi^+]_8^{(4)}\!\!&=&\!\!-\frac{8M^2}{f^2}
\left[(2\beta^8_1+2\beta^8_3+2\beta^8_{10}+\beta^8_{11})M^2+
\beta^8_2 NM_{SS}^2\right. \label{KPIOCTETCT} \\
&&\hspace{-1.6truecm}\left.
+16\alpha^8_1((\lambda_4-\lambda_6)NM_{SS}^2+(\lambda_5-\lambda_8)M^2)
-8\alpha^8_2(\lambda_4 NM_{SS}^2+\lambda_5 M^2)\right]\;, \nonumber \\
(4\pi f)^2[K^+\to\pi^+]_{27}^{(4)}\!\!&=&\!\!-\frac{8M^2}{f^2}
\left[2(\beta^{27}_2+\beta^{27}_4)M^2+\beta^{27}_7 NM_{SS}^2\right.
\label{KPITSPLETCT} \\
&&\phantom{\!\!-\frac{8M^2}{f^2}}\left.
-16\alpha^{27}((\lambda_4-\lambda_6)NM_{SS}^2+
(\lambda_5-\lambda_8)M^2)\right]\;, \nonumber
\ea
with again the $\Delta I=1/2$ and $\Delta I=3/2$ results the same
for the 27-plet.  The $\lambda_i$ are the strong $O(p^4)$ LECs;
they are related to the Gasser--Leutwyler $L_i$ \cite{gasleu} by
\be
\lambda_i=16\pi^2 L_i\;. \label{LI}
\ee
We see an example here of the fact that a 
partially quenched simulation, with $M\ne M_{SS}$, would in principle 
yield more information about the $O(p^4)$ LECs than an unquenched
simulation in which $M=M_{SS}$.

For the $K\to 0$ matrix elements we take non-degenerate valence quarks
with $m_s\neq m_d$ and $m_d =m_u$, and define
\be
M_{33}^2=2M_K^2-M_\pi^2\;. \label{MSTRANGE}
\ee
The pion will be made out of two
light valence quarks, and the kaon out of a light and a strange
valence quark.  We also define $M_{iS}^2$ to be the (tree-level)
mass of a meson
made out of the $i$-th valence quark and a sea quark, 
\be
M_{iS}^2=B_0(m_i+m_S)\;,\ \ \ i=u,d,s\;. \label{MIS}
\ee
Of course, $M_{uS}=M_{dS}$.
For the octet matrix element we find, at one loop,
\ba
&&\hspace{-0.7truecm}[K\to 0]_8=-\frac{4i\alpha^8_1}{f(4\pi f)^2}
\Biggl[N(M_{uS}^4(L(M_{uS})-1)
-M_{sS}^4(L(M_{sS})-1)) \label{KVAOCTET} \\
&&\hspace{3.6truecm}
+\frac{2}{3}\cM^2\left(M_\pi^2(L(M_\pi)-\frac{1}{2})
-M_{33}^2(L(M_{33})-\frac{1}{2})\right) \nonumber \\
&&\hspace{3.55truecm}
-A\left(M_\pi^4(L(M_\pi)-\frac{2}{3})
-M_{33}^4(L(M_{33})-\frac{2}{3})\right)\Biggr] \nonumber \\
&&\hspace{0.1truecm}+\frac{4i\alpha^8_2(M_K^2-M_\pi^2)}{f}
\left(1+\frac{1}{(4\pi f)^2}\left[
-\frac{N}{2}\left(M_{uS}^2(L(M_{uS})-1)+M_{sS}^2(L(M_{sS})-1)\right)
\right.\right. \nonumber \\
&&\hspace{2.55truecm}
-\frac{1}{6}\cM^2\left(L(M_\pi)+L(M_{33})+2L_1(M_{33},M_\pi)-2\right)
\nonumber \\
&&\hspace{2.0truecm}\left.\left.
+\frac{1}{6}A\left(2M_\pi^2L(M_\pi)+2M_{33}^2L(M_{33})
+2L_2(M_{33},M_\pi)-3M_\pi^2-3M_{33}^2\right)\right]\right) \nonumber \\
&&\hspace{0.1truecm}
+\frac{2i\gamma^8_1}{f(4\pi f)^2}\Biggl[
M_\pi^4(L(M_\pi)-1)-M_{33}^4(L(M_{33})-1) \nonumber \\
&&\hspace{2.2truecm}
+\frac{N}{3}(\cM^2-AM_\pi^2)\left(L_2(M_{SS},M_\pi)-M_{SS}^2-M_\pi^2\right)
\nonumber \\
&&\hspace{2.2truecm} 
-\frac{N}{3}(\cM^2-AM_{33}^2)\left(L_2(M_{SS},M_{33})-M_{SS}^2-M_{33}^2\right)
\Biggr] \nonumber \\
&&\hspace{-0.4truecm}
-\frac{4i\gamma^8_2(M_K^2-M_\pi^2)}{f(4\pi f)^2}\left[
M_\pi^2(L(M_\pi)-1)+M_{33}^2(L(M_{33})-1)
-\frac{2N}{3}AM_{SS}^2(L(M_{SS})-1)\right. \nonumber \\
&&\hspace{0.45truecm} \left.
+\frac{N}{3}(\cM^2-AM_\pi^2)(L_1(M_{SS},M_\pi)-1)
+\frac{N}{3}(\cM^2-AM_{33}^2)(L_1(M_{SS},M_{33})-1)\right]. \nonumber
\ea
In this case, the leading order $p^2$ contribution comes
 only from the weak 
mass term proportional to $\alpha_2^8$.  
 Unlike the case of $K\to\pi$,
the contributions from the $O(p^2)$ $\eta'-$operators proportional
to $\gamma^8_1$ and $\gamma^8_2$ are different in this case.
The argument explaining the situation in the $K\to\pi$ case
does not work here, because the total derivative in the partial
integration cannot be dropped, as the process $K\to 0$ does
not conserve energy.

For the 27-plet (both $\Delta I=1/2$ and $\Delta I=3/2$), we obtain
\ba
&&\hspace{-0.7truecm}[K\to 0]_{27}=
\frac{12i\alpha^{27}}{f(4\pi f)^2}\Biggl[
M_\pi^4(L(M_\pi)-1)+M_{33}^4(L(M_{33})-1)-2M_K^4(L(M_K)-1)
\nonumber \\
&&\hspace{1.35truecm}
+\frac{2}{3}\cM^2\left(M_\pi^2L(M_\pi)+M_{33}^2L(M_{33})
-L_2(M_{33}, M_\pi)+\frac{1}{2}(M_\pi^2+M_{33}^2)\right) \nonumber \\
&&\hspace{1.3truecm}
-\frac{1}{3}A\left(L_3(M_{33},M_\pi)-3M_\pi^2M_{33}^2
L_1(M_{33},M_\pi)+2M_\pi^2 M_{33}^2\right)\Biggr]\,.
\label{KVATSPLET}
\ea
Finally, the $O(p^4)$ operators of Eqs.~(\ref{OCTET},\ref{TSPLET},%
\ref{OCTETTOTDER}) give
\ba
(4\pi f)^2[K\to 0]_8^{(4)}\!&=&\!
\frac{8i(M_K^2-M_\pi^2)}{f}\left[
(2\beta^8_1-2\beta^8_5+\tb^8_1)M_K^2+\beta^8_2 NM_{SS}^2 \right. 
\nonumber \\
&&\hspace{3truecm}\left.
-4\alpha^8_2\left(\lambda_4 NM_{SS}^2
+\lambda_5 M_K^2
\right)\right]\;, \label{KVAOCTETCT} \\
(4\pi f)^2[K\to 0]_{27}^{(4)}\!&=&\!
-\frac{12i}{f}\,4\beta^{27}_1(M_K^2-M_\pi^2)^2\;. \label{KVATSPLETCT}
\ea
These results hold for an arbitrary number $N$ of degenerate sea
quarks.  Note the
appearance of ``quenched chiral logs," contained in the one-loop
logarithms proportional to $\cM^2$.  Since $\cM^2$ does not depend on
the valence masses, such terms decrease with decreasing valence
quark masses at the same rate as tree-level terms (modulo the
logarithms), {\it i.e.} typically as $m\log{m}$ instead of
$m^2\log{m}$.

Our results are presented here in a form somewhat different from that  
in Ref.~\cite{mgeplat99}.  Here, we express the matrix elements in terms
of the tree-level meson masses, or equivalently, the quark masses,
while in Ref.~\cite{mgeplat99} they were expressed in terms of 
renormalized masses (also, only the chiral logarithms were given).

These results can be converted into expressions for the matrix
elements as a function of the actual meson masses computed on
the lattice by using the one-loop expression for the  mass of a
meson made out of two non-degenerate valence quarks in terms of the
tree-level masses, Eq.~(\ref{MMASS}).  This expression, in 
$\overline{MS}$, and including $O(p^4)$ contributions, is
\ba
\left(M_K^2\right)^{\rm 1-loop}&=&
M_K^2\left(1+\frac{1}{(4\pi f)^2}\Biggl[
-\frac{2}{3}\cM^2\left(L_1(M_{33},M_\pi)-1\right)\right.
\label{OLMASS} \\
&&\phantom{M_K^2\Bigl(1+\frac{1}{(4\pi f)^2}\Bigl[}
+\frac{2}{3}A\left(L_2(M_{33},M_\pi)-M_{33}^2-M_\pi^2\right)
\nonumber \\
&&\phantom{M_K^2\Bigl(1+\frac{1}{(4\pi f)^2}\Bigl[}\left.
+16\left(M_K^2(2\lambda_8-\lambda_5)+NM_{SS}^2(2\lambda_6
-\lambda_4)\right)\Biggr]\right)\;. \nonumber
\ea
For degenerate valence-quark masses, this simplifies to
\ba
\left(M_\pi^2\right)^{\rm 1-loop}&=&
M^2\left(1+\frac{1}{(4\pi f)^2}\Biggl[
-\frac{2}{3}\cM^2L(M)
+\frac{2}{3}AM^2\left(2L(M)-1\right)\right.
\label{OLMASSDEG} \\
&&\phantom{M_K^2\Bigl(1+\frac{1}{(4\pi f)^2}\Bigl[}\left.
+16\left(M^2(2\lambda_8-\lambda_5)+NM_{SS}^2(2\lambda_6
-\lambda_4)\right)\Biggr]\right)\;. \nonumber
\ea

\subsection{\bf Partially quenched results for large $M_{\eta'}$}

The results presented above simplify when 
$M_{\eta'}$ is taken large compared to both the sea- and valence-%
meson masses. 
Taking $M_{\eta'}$ large while keeping $M_{ii,jj}$ and
$M_{SS}$ fixed in Eq.~(\ref{X}) gives
\be
X_{ij}(p)=\frac{1}{N}\left(\frac{1}{p^2+M^2_{ii}}
+\frac{M^2_{SS}-M^2_{jj}}
{(p^2+M^2_{ii})(p^2+M^2_{jj})}\right)
\;, \label{SLIMIT}
\ee
dropping the $\eta'$ pole.  This expression for $X_{ij}(p)$ leads
to the simplified expressions in Eq.~(\ref{CALMLIMIT}) for
$\cM^2$ and $A$, where all the $\eta'$ parameters have disappeared, and the 
$\eta'$ meson has been decoupled.  Thus, for $K\to \pi$, we obtain
\ba
[K^+\to\pi^+]_8&=&\frac{4\alpha^8_1M^2}{f^2}\left(1-
\frac{1}{(4\pi f)^2}\Biggl[N\Mbar^2(L(\Mbar)-1)
\right.
\label{SKPIOCTET} \\
&&\phantom{\frac{4\alpha^8_1M^2}{f^2}\Bigl(} \left.
+\frac{2}{N}\left((3M_{SS}^2-8M^2)L(M)
+M_{SS}^2+2M^2\right)\Biggr]\right) \nonumber \\
&&
-\,\frac{4\alpha^8_2M^2}{f^2}\left(1-\frac{1}{(4\pi f)^2}
\frac{2}{N}\left[
(M_{SS}^2-4M^2)L(M)+M_{SS}^2\right]\right)\;,
\nonumber 
\ea
and 
\ba
[K^+\to\pi^+]_{27}&\!\!=&\!\!-\frac{4\alpha^{27}M^2}{f^2}\left(1-
\frac{1}{(4\pi f)^2}\Biggl[2N\Mbar^2(L(\Mbar)-1) 
\right.  \label{SKPITSPLET} \\
&&\phantom{-\frac{\alpha^{27}M}{f^2}}
\left.+\,2M^2(3L(M)-2)
+\frac{2}{N}\left((M_{SS}^2-2M^2)L(M)+M^2\right)
\Biggr]\right)\, , \nonumber
\ea
while for $K\to 0$, we find
\ba
&&\hspace{-0.7truecm}[K\to 0]_8=-\frac{4i\alpha^8_1}{f(4\pi f)^2}
\Biggl[N(M_{uS}^4(L(M_{uS})-1)
-M_{sS}^4(L(M_{sS})-1) \label{SKVACOCTET} \\
&&\hspace{3.6truecm}
+\frac{2}{N}M_{SS}^2\left(M_\pi^2(L(M_\pi)-\frac{1}{2})
-M_{33}^2(L(M_{33})-\frac{1}{2})\right) \nonumber \\
&&\hspace{3.55truecm}
-\frac{3}{N}\left(M_\pi^4(L(M_\pi)-\frac{2}{3})
-M_{33}^4(L(M_{33})-\frac{2}{3})\Biggr)\right] \nonumber \\
&&\hspace{0.1truecm}+\frac{4i\alpha^8_2(M_K^2-M_\pi^2)}{f}
\left(1+\frac{1}{(4\pi f)^2}\left[
-\frac{N}{2}\left(M_{uS}^2(L(M_{uS})-1)+M_{sS}^2(L(M_{sS})-1)\right)
\right.\right. \nonumber \\
&&\hspace{1.95truecm}
-\frac{1}{2N}M_{SS}^2\left(L(M_\pi)+L(M_{33})+2L_1(M_{33},M_\pi)-2\right)
\nonumber \\
&&\hspace{1.9truecm}\left.\left.
+\frac{1}{2N}\left(2M_\pi^2L(M_\pi)+2M_{33}^2L(M_{33})
+2L_2(M_{33},M_\pi)-3M_\pi^2-3M_{33}^2\right)\right]\right) \nonumber
\ea
for the octet, and
\ba
&&\hspace{-0.7truecm}[K\to 0]_{27}=
\frac{12i\alpha^{27}}{f(4\pi f)^2}\Biggl[
M_\pi^4(L(M_\pi)-1)+M_{33}^4(L(M_{33})-1)-2M_K^4(L(M_K)-1)
\nonumber \\
&&\hspace{1.35truecm}
+\frac{2}{N}M_{SS}^2\left(M_\pi^2L(M_\pi)+M_{33}^2L(M_{33})
-L_2(M_{33}, M_\pi)+\frac{1}{2}(M_\pi^2+M_{33}^2)\right) \nonumber \\
&&\hspace{1.3truecm}
-\frac{1}{N}\left(L_3(M_{33},M_\pi)-3M_\pi^2M_{33}^2
L_1(M_{33},M_\pi)+2M_\pi^2 M_{33}^2\right)\Biggr]
\label{SKVATSPLET}
\ea
for the 27-plet.  All the dependence on $\eta'$ parameters
($m_0^2$, $\alpha$, $\gamma^8_{1,2}$) has disappeared from these
expressions, as expected. The contributions from $O(p^4)$
operators, given in Eqs.~(\ref{KPIOCTETCT},\ref{KPITSPLETCT})
and (\ref{KVAOCTETCT},\ref{KVATSPLETCT}), do not change.
Note however, as mentioned before, that only in this limit the
corresponding LECs are independent of the quark masses. 

\subsection{Quenched results}

A special case of practical interest is the completely quenched
result.  In the quenched approximation, there are no sea quarks,
and hence, quenched expressions can be obtained by setting $N=0$
in Eqs.~(\ref{KPIOCTET}--\ref{KPITSPLETCT},\ref{KVAOCTET},%
\ref{KVATSPLET},\ref{KVAOCTETCT},\ref{KVATSPLETCT}), or equivalently,
by taking $M_{SS}\to\infty$.  In this
case, it is not possible to decouple the $\eta'$ 
\cite{sharpecapri,bgq}.
The expressions given below maybe rewritten in terms of the
parameter $\delta$ (introduced in Ref.~\cite{sharpeq}), by 
setting
\be
m_0^2=24\pi^2 f^2\delta\;. \label{DELTA} 
\ee
The quenched results are
\ba
&&\hspace{-0.7truecm}[K^+\to\pi^+]_8=\frac{4\alpha^8_1M^2}{f^2}\left(1-
\frac{1}{(4\pi f)^2}\left[
2\left(m_0^2-\frac{8}{3}\alpha M^2\right)L(M)
+\frac{2}{3}(m_0^2+2\alpha M^2)\right]\right) \nonumber \\
&&\hspace{1.6truecm}
-\,\frac{4\alpha^8_2M^2}{f^2}\left(1-\frac{1}{(4\pi f)^2}\left[
\frac{2}{3}(m_0^2-
4\alpha M^2)L(M)+\frac{2}{3}m_0^2\right]\right)
\label{QKPIOCTET} \\
&&\hspace{1.6truecm}
+\frac{4(\gamma^8_1+2\gamma^8_2)M^2}{f^2(4\pi f)^2}\;
\left[2M^2L(M)-M^2\right]\;,
\nonumber
\ea
\ba
&&\hspace{-4.1truecm}[K^+\to\pi^+]_{27}=
-\frac{4\alpha^{27}M^2}{f^2}\left(\!1-\frac{1}{(4\pi f)^2}
\Biggl[2M^2(3L(M)-2)\right. \label{QKPITSPLET} \\
&&\hspace{1truecm}
\left.+\frac{2}{3}(m_0^2-2\alpha M^2)L(M)+\frac{2}{3}
\alpha M^2 \Biggr] \right)\;, \nonumber
\ea
\ba
&&\hspace{-0.7truecm}[K\to 0]_8=-\frac{4i\alpha^8_1}{f(4\pi f)^2}
\left[
\frac{2}{3}m_0^2\left(M_\pi^2(L(M_\pi)-\frac{1}{2})
-M_{33}^2(L(M_{33})-\frac{1}{2})\right)\right. \label{QKVAOCTET} \\
&&\hspace{3.55truecm}\left.
-\alpha\left(M_\pi^4(L(M_\pi)-\frac{2}{3})
-M_{33}^4(L(M_{33})-\frac{2}{3})\right)\right] \nonumber \\
&&\hspace{0.1truecm}+\frac{4i\alpha^8_2(M_K^2-M_\pi^2)}{f}
\left(1+\frac{1}{(4\pi f)^2}\left[
-\frac{1}{6}m_0^2\left(L(M_\pi)+L(M_{33})+2L_1(M_{33},M_\pi)-2\right)
\right.\right. \nonumber \\
&&\hspace{2.0truecm}\left.\left.
+\frac{1}{6}\alpha\left(2M_\pi^2L(M_\pi)+2M_{33}^2L(M_{33})
+2L_2(M_{33},M_\pi)-3M_\pi^2-3M_{33}^2\right)\right]\right) \nonumber
\\
&&\hspace{0.1truecm}
+\frac{2i\gamma^8_1}{f(4\pi f)^2}\left[
M_\pi^4(L(M_\pi)-1)-M_{33}^4(L(M_{33})-1)\right] \nonumber \\
&&\hspace{0.1truecm}
-\frac{4i\gamma^8_2(M_K^2-M_\pi^2)}{f(4\pi f)^2}\left[
M_\pi^2(L(M_\pi)-1)+M_{33}^2(L(M_{33})-1)
\right]\;, \nonumber 
\ea
\ba
&&\hspace{-0.7truecm}[K\to 0]_{27}=
\frac{12i\alpha^{27}}{f(4\pi f)^2}\Biggl[
M_\pi^4(L(M_\pi)-1)+M_{33}^4(L(M_{33})-1)-2M_K^4(L(M_K)-1)
\nonumber \\
&&\hspace{1.35truecm}
+\frac{2}{3}m_0^2\left(M_\pi^2L(M_\pi)+M_{33}^2L(M_{33})
-L_2(M_{33}, M_\pi)+\frac{1}{2}(M_\pi^2
+M_{33}^2)\right) \nonumber \\
&&\hspace{1.3truecm}
-\frac{1}{3}\alpha\left(L_3(M_{33},M_\pi)+3M_\pi^2M_{33}^2
L_1(M_{33},M_\pi)+2M_\pi^2 M_{33}^2\right)\Biggr]\,.
\label{QKVATSPLET}
\ea
Contributions from $O(p^4)$ operators are obtained by
setting $N=0$ in Eqs.~(\ref{KPIOCTETCT},\ref{KPITSPLETCT}) and
(\ref{KVAOCTETCT},\ref{KVATSPLETCT}).  However, we emphasize
 that all the information
obtained from quenched lattice computations is about the $N=0$ values
of the LECs appearing in these equations.  These values are,
in principle, different from their $N=3$ values. 

\subsection{Unquenched results}

For completeness, we also report the results for the unquenched theory
with three light flavors. 
For $K\to\pi$ these can be simply obtained by setting $N=3$ and
$M_{SS}^2=M^2$ in our (large-$M_{\eta'}$)
partially quenched expressions of Subsec.~4.2.  For
$K\to 0$ we have to choose the sea-quark masses 
equal to the non-degenerate
valence-quark masses.  The results for $K\to 0$ therefore cannot
be derived from our
partially quenched results, where we took all sea quarks to
be degenerate in mass from the start.  The results are
\ba
[K^+\to\pi^+]_8&=&\frac{4\alpha^8_1 M^2}{f^2}\left(
1+\frac{1}{(4\pi f)^2}\left[\frac{1}{3}M^2L(M)+M^2\right]\right)
\label{UKPIOCTET} \\
&& -\frac{4\alpha^8_2 M^2}{f^2}\left(
1+\frac{1}{(4\pi f)^2}\left[2M^2L(M)-\frac{2}{3}M^2\right]\right)\,,
\nonumber \\
{}[K^+\to\pi^+]_{27}&=&-\frac{4\alpha^{27} M^2}{f^2}\left(
1-\frac{1}{(4\pi f)^2}\left[\frac{34}{3}M^2L(M)
-\frac{28}{3}M^2\right]\right) 
\label{UKPITSPLET}
\ea 
for $K\to\pi$, and
\ba
[K\to 0]_8\!\!&=&\!\!\frac{4i\alpha^8_2}{f}(M_K^2-M_\pi^2)
\left(1+\frac{1}{(4\pi f)^2}\left[
-\frac{3}{4}M_\pi^2L(M_\pi)-\frac{3}{2}M_K^2L(M_K) \right.\right.
\label{UKVAOCTET} \\
&&\hspace{3.3truecm}\left.\left.
-\frac{1}{12}M_\eta^2L(M_\eta)
+\frac{29}{18}M_K^2+\frac{13}{18}M_\pi^2\right]\right)
\nonumber \\
&&\!+\frac{4i\alpha^8_1}{f}\,\frac{M_K^2-M_\pi^2}{(4\pi f)^2}
\left[\frac{1}{3}L_2(M_\eta,M_K)-2L_2(M_\eta,M_\pi)
+\frac{17}{9}M_K^2+\frac{13}{9}M_\pi^2\right]\,,
\nonumber \\
{}[K\to 0]_{27}\!\!&=&\!\!
\frac{4i\alpha^{27}}{f}\,\frac{M_K^2-M_\pi^2}{(4\pi f)^2}
\left[-2L_2(M_\eta,M_K)+2L_2(M_\eta,M_\pi)+2M_K^2-2M_\pi^2\right]
\nonumber \\
\ea
for $K\to 0$.  The contributions from $O(p^4)$ operators are
obtained from the partially quenched expressions
Eqs.~(\ref{KPIOCTETCT},\ref{KPITSPLETCT},\ref{KVAOCTETCT},%
\ref{KVATSPLETCT}), by replacing $NM_{SS}^2\to 2M_K^2+M_\pi^2$
in those equations.
\section{Relation to $K\to\pi\pi$ and numerical examples}
\secteq{5}
We now turn to a discussion on how our results can be used
to extract physical information from lattice results.  

If tree-level ChPT were a good approximation, one could
determine $\alpha^8_{1,2}$ and $\alpha^{27}$ from a lattice
computation of the $K\to\pi$ and $K\to 0$ matrix elements, and
then use ChPT to predict the $\Delta I=1/2$ and $\Delta I=3/2$
$K\to\pi\pi$ decay rates \cite{bernardetal}.  
For instance, for the $\Delta I=1/2$
matrix element, one finds
\ba
[K^0\to\pi^+\pi^-]_{1/2}&=&\frac{4i(m_K^2-m_\pi^2)}{f^3}
(\alpha^8_1-\alpha^{27}) \label{TREEEX} \\
&=&\frac{i}{f}\frac{m_K^2-m_\pi^2}{M^2}
\left([K^+\to\pi^+]_{1/2}-b[K^0\to 0]\right)\;,\nonumber \\
b&\equiv&\frac{iM^2}{f(M_K^2-M_\pi^2)}=
\frac{2im}{f(m_s-m_d)}\;. \label{B} 
\ea
Here $m_K$ and $m_\pi$ are the physical kaon and pion masses, 
$M$ is the degenerate meson mass 
(corresponding to a degenerate quark mass $m$)
used in the lattice computation
of $[K^+\to\pi^+]_{1/2}$, and $M_K$ and $M_\pi$ are the
non-degenerate meson masses (corresponding to quark masses $m_s$ and
$m_d$) used in the lattice computation of
$[K^0\to 0]$.  At tree level, the procedure is very simple,
because the conversion involves only meson masses and the
meson decay constant $f$, which can relatively easily be
determined.

To repeat a similar analysis at one loop, one would not only
need to eliminate the $O(p^2)$ constants $\alpha^8_{1,2}$ and
$\alpha^{27}$, but also all the $O(p^4)$ LECs that can appear
in the matrix elements for the kaon decays of interest.  In other
words, the $O(p^4)$ weak
LECs $\beta^8_{1,2,3,10,11,13,15}$ and $\beta^{27}_{1,2,4,5,6,7}$
are needed \cite{bpp}, as well as the strong LECs
$\lambda_{4,5,6,8}$.  Only a few linear combinations
of those can be determined from $K\to\pi$ and $K\to 0$ matrix
elements.  For
the $K\to\pi\pi$ matrix elements also only a few linear
combinations are needed, but these are different linear combinations,
involving also $\beta^8_{13,15}$ and $\beta^{27}_{5,6}$ which do not 
even appear in $K\to\pi$ and $K\to 0$ at all.  

Also, we have seen that in
unphysical matrix elements like $K\to 0$ new LECs appear in these
linear combinations, such as for instance $\tb^8_1$ in the
linear combination
$\beta^8_1-2\beta^8_5+\tb^8_1$ in Eq.~(\ref{KVAOCTETCT}).  
Likewise, one would be able to determine more LECs from $K\to\pi$
and $K\to\eta$ in the mass non-degenerate case, but also more
unphysical (total-derivative) $O(p^4)$ operators would
contribute, since these matrix elements would also not
conserve energy for onshell external states, just as $[K^0\to 0]$.

The only other way to determine more of the LECs from a lattice
computation would be to consider more complicated correlation
functions (such as $K\to\pi\pi$ itself).  In that
case, one necessarily has more than one strongly interacting
particle in the initial or final states, and this 
leads to rather severe complications of its own (for
$K\to\pi\pi$, see Refs.~\cite{mgeplat99,mgep}).  In general,
on the lattice, one only has access to these matrix elements
for unphysical choices of the kinematics \cite{mt,gl,dawsonetal}.
Again, not all relevant $O(p^4)$ LECs can be determined.
 
We conclude that at one loop uncertainties are introduced in
the determination of $K\to\pi\pi$ matrix elements from
$K\to\pi$ and $K\to 0$, which are not present at tree level.
These uncertainties break down into two parts.  One is the
determination of $\alpha^8_{1,2}$, $\alpha^{27}$ from
$K\to\pi$ and $K\to 0$, and the other is the conversion of
results for these $O(p^2)$ LECs into $K\to\pi\pi$ decay
rates.  Here, we will only consider the first part,
{\it i.e.} we will concentrate on the determination
of $\alpha^8_{1,2}$ and $\alpha^{27}$ from $K\to\pi$ and
$K\to 0$ matrix elements.  

If we are only interested in the
determination of $\alpha^8_{1,2}$ and $\alpha^{27}$, we need
to know only the polynomial form (in the meson masses)
of the contributions from $O(p^4)$ operators, as given in
Eqs.~(\ref{KPIOCTETCT},\ref{KPITSPLETCT},\ref{KVAOCTETCT},%
\ref{KVATSPLETCT}),  assuming that one-loop ChPT can be
reliably applied to lattice computations of $K\to\pi$ and
$K\to 0$ matrix elements. 
The results of the previous section can be written in the
generic form
\ba
[K\to\pi]_8&=&\frac{4M^2}{f^2}\left[\alpha^8_1\left(1+X^8_1\right)
-\alpha^8_2\left(1+X^8_2\right)+(\gamma^8_1+2\gamma^8_2)X^\gamma
\right. \label{GENERIC} \\
&&\phantom{\frac{4M^2}{f^2}\Bigl[\alpha^8_1\left(1+X^8_1\right)
-\alpha^8_2\left(1+X^8_2\right)}\left.
+C^8_V\frac{M^2}{(4\pi f)^2}+C^8_S\frac{NM_{SS}^2}{(4\pi f)^2}\right]\;, 
\nonumber \\
{}[K\to\pi]_{27}&=&-\frac{4M^2}{f^2}\left[\alpha^{27}
\left(1+X^{27}\right)+C^{27}_V\frac{M^2}{(4\pi f)^2}
+C^{27}_S\frac{NM_{SS}^2}{(4\pi f)^2}\right]\;,
\nonumber \\
{}[K\to 0]_8&=&\frac{4i(M_K^2-M_\pi^2)}{f}\left[\alpha^8_1 Y^8_1
+\alpha^8_2\left(1+Y^8_2\right)+\gamma^8_1 Y^\gamma_1+\gamma^8_2 Y^\gamma_2
\right. \nonumber \\
&&\phantom{\frac{4i(M_K^2-M_\pi^2)}{f}\Bigl[\alpha^8_1 Y^8_1
+\alpha^8_2\left(1+Y^8_2\right)}\left.
+D^8_V\frac{M_K^2}{(4\pi f)^2}+D^8_S\frac{NM_{SS}^2}{(4\pi f)^2}\right]\;,
\nonumber \\
{}[K\to 0]_{27}&=&\frac{4i(M_K^2-M_\pi^2)}{f}\left[\alpha^{27}Y^{27}
+D^{27}\frac{M_K^2-M_\pi^2}{(4\pi f)^2}\right]\;.
\nonumber
\ea
In these equations, $X^8_{1,2}$, $X^{27}$, $Y^8_{1,2}$, $Y^{27}$,
$X^\gamma$ and $Y^\gamma_{1,2}$
stand for the one-loop contributions (``chiral logarithms")
given explicitly in 
Eqs.~(\ref{KPIOCTET},\ref{KPITSPLET},\ref{KVAOCTET},\ref{KVATSPLET}).
The constants $C^8_{V,S}$ {\it etc.} are
linear combinations of $O(p^4)$ LECs, which can be expressed
in terms of the $\beta$'s and $\lambda$'s by comparison with
Eqs.~(\ref{KPIOCTETCT},\ref{KPITSPLETCT},\ref{KVAOCTETCT},%
\ref{KVATSPLETCT}).
{}From these relations $\alpha^8_{1,2}$ and $\alpha_{27}$ can be
extracted by fitting these equations to lattice results for
the matrix elements at various different values of the quark masses.
With sufficient precision, also the $O(p^4)$ LECs could in
principle be determined, but it is unlikely that this will
work in practice with the currently available computational
power.  However, this does not imply that the $O(p^2)$ LECs
cannot be determined with a reasonable accuracy.

In order to get an idea about the size of one-loop effects, we
will set the $O(p^4)$ coefficients $C^{8,27}_{V,S}$,
$D^8_{V,S}$ and $D^{27}$ to zero, 
and evaluate the chiral logarithms 
at typical lattice values of the parameters and at $\Lambda=1$~GeV,
$\Lambda=m_\rho=770$~MeV, and $\Lambda=m_\eta=550$~MeV.  We will
consider three different ``theories," partially quenched with
$N=2$ or $3$ and $M_\eta'$ large, and quenched
($N=0$) with arbitrary $\delta$. We will 
also set $f=f_\pi=132$~MeV.  We take $M_{SS}=500$~MeV,
which corresponds to a sea-quark mass of about half the
strange quark mass,  vary the degenerate ``lattice"
meson mass $M$ at which $[K\to\pi]$
is determined, and take $2M_K^2/3=M_\pi^2=M^2$ for $[K\to 0]$,
which corresponds to $m_s=2m_d=2m$. 

\begin{table}[t]
\begin{center}
\begin{tabular}{|c|c|c|c|c|} \hline
$N$ & $M$ & $X^8_1$ & $X^8_2$ &
$\hX^{27}$ \\ \hline
3 & 200 & 0.72 &  0.01 & 0.34 \\ \hline
3 & 350 & 0.31 & -0.18 & 0.74 \\ \hline
3 & 500 & 0.05 & -0.31 & 1.12 \\ \hline
2 & 200 & 0.69 &  0.01 & 0.34 \\ \hline
2 & 350 & 0.01 & -0.27 & 0.74 \\ \hline
2 & 500 & -0.47 & -0.47 & 1.12 \\ \hline
0 & 200 & 0.87$(\delta/0.1)$ & 0.22$(\delta/0.1)$ &
0.34 \\ \hline
0 & 350 & 0.53$(\delta/0.1)$ & 0.11$(\delta/0.1)$ &
0.74 \\ \hline
0 & 500 & 0.32$(\delta/0.1)$ & 0.04$(\delta/0.1)$ &
1.12 \\ \hline
\end{tabular}
\end{center}
\caption{The one-loop corrections $X^8_{1,2}$ and $X^{27}$ for
$2M_K^2/3=M_\pi^2=M^2$, $M_{SS}=500$~MeV, $\Lambda=1$~GeV,
$M$ in MeV.  The $N=2,3$ values
have been calculated for large $M_{\eta'}$, {\it i.e.}
using the results of Subsect~4.2.} 
\vspace{0.2cm}
\label{tb:xexamples1}
\end{table}
\begin{table}[t]
\begin{center}
\begin{tabular}{|c|c|c|c|c|} \hline
$N$ & $M$ & $X^8_1$ & $X^8_2$ &
$\hX^{27}$ \\ \hline
3 & 200 & 0.58 & 0.00 & 0.29 \\ \hline
3 & 350 & 0.23 & -0.15 & 0.60 \\ \hline
3 & 500 & 0.06 & -0.22 & 0.83 \\ \hline
2 & 200 & 0.56 & 0.00 & 0.29 \\ \hline
2 & 350 & -0.02 & -0.23 & 0.60 \\ \hline
2 & 500 & -0.33 & -0.33 & 0.83 \\ \hline
0 & 200 & 0.71$(\delta/0.1)$ & 0.17$(\delta/0.1)$ & 
0.29 \\ \hline
0 & 350 & 0.37$(\delta/0.1)$ & 0.06$(\delta/0.1)$ &
0.60 \\ \hline
0 & 500 & 0.16$(\delta/0.1)$ & -0.01$(\delta/0.1)$ &
0.83 \\ \hline
\end{tabular}
\end{center}
\caption{As in Table~\ref{tb:xexamples1}, but with $\Lambda=770$~MeV.}
\vspace{0.2cm}
\label{tb:xexamples2}
\end{table}
\begin{table}[t]
\begin{center}
\begin{tabular}{|c|c|c|c|c|} \hline
$N$ & $M$ & $X^8_1$ & $X^8_2$ &
$\hX^{27}$ \\ \hline
3 & 200 & 0.41 & -0.02 & 0.23 \\ \hline 
3 & 350 & 0.13 & -0.11 & 0.42 \\ \hline
3 & 500 & 0.09 & -0.10 & 0.47 \\ \hline
2 & 200 & 0.38 & -0.02 & 0.23 \\ \hline
2 & 350 & -0.05 & -0.17 & 0.42 \\ \hline
2 & 500 & -0.14 & -0.14 & 0.47 \\ \hline
0 & 200 & 0.51$(\delta/0.1)$ & 0.10$(\delta/0.1)$ & 
0.23 \\ \hline
0 & 350 & 0.17$(\delta/0.1)$ & -0.01$(\delta/0.1)$ &
0.42 \\ \hline
0 & 500 & -0.04$(\delta/0.1)$ & -0.08$(\delta/0.1)$ &
0.47 \\ \hline
\end{tabular}
\end{center}
\caption{As in Table~\ref{tb:xexamples1}, but with $\Lambda=550$~MeV.}
\vspace{0.2cm}
\label{tb:xexamples3}
\end{table}
\begin{table}[t]
\begin{center}
\begin{tabular}{|c|c|c|c|c|} \hline
$N$ & $M$ & $Y^8_1$ & $Y^8_2$ &
$Y^{27}$ \\ \hline
3 & 200 & -0.82 & 0.56 & 0.00 \\ \hline
3 & 350 & -0.61 & 0.51 & -0.04 \\ \hline
3 & 500 & -0.57 & 0.53 & -0.06 \\ \hline
2 & 200 & -0.57 & 0.44 & 0.05 \\ \hline
2 & 350 & -0.12 & 0.27 & 0.02 \\ \hline
2 & 500 & 0.08 & 0.19 & -0.03 \\ \hline
0 & 200 & -0.47$(\delta/0.1)$ & 0.29$(\delta/0.1)$ & 
-0.10+0.07$(\delta/0.1)$ \\ \hline
0 & 350 & -0.24$(\delta/0.1)$ & 0.17$(\delta/0.1)$ &
-0.16+0.07$(\delta/0.1)$ \\ \hline
0 & 500 & -0.10$(\delta/0.1)$ & 0.10$(\delta/0.1)$ &
-0.13+0.07$(\delta/0.1)$ \\ \hline
\end{tabular}
\end{center}
\caption{The one-loop corrections $Y^8_{1,2}$ and $Y^{27}$ for
$2M_K^2/3=M_\pi^2=M^2$, $M_{SS}=500$~MeV, $\Lambda=1$~GeV,
$M$ in MeV. The $N=2,3$ values have been calculated 
for large $M_{\eta'}$, {\it i.e.} using the results of Subsect~4.2.} 
\vspace{0.2cm}
\label{tb:yexamples1}
\end{table}
\begin{table}[t]
\begin{center}
\begin{tabular}{|c|c|c|c|c|} \hline
$N$ & $M$ & $Y^8_1$ & $Y^8_2$ &
$Y^{27}$ \\ \hline
3 & 200 & -0.63 & 0.46 & 0.01 \\ \hline
3 & 350 & -0.44 & 0.40 & -0.02 \\ \hline
3 & 500 & -0.44 & 0.41 & -0.02 \\ \hline
2 & 200 & -0.42 & 0.36 & 0.05 \\ \hline
2 & 350 & -0.07 & 0.21 & 0.02 \\ \hline
2 & 500 & -0.02 & 0.16 & -0.03 \\ \hline
0 & 200 & -0.36$(\delta/0.1)$ & 0.23$(\delta/0.1)$ & 
-0.08+0.07$(\delta/0.1)$ \\ \hline
0 & 350 & -0.14$(\delta/0.1)$ & 0.12$(\delta/0.1)$ &
-0.09+0.07$(\delta/0.1)$ \\ \hline
0 & 500 & 0.00$(\delta/0.1)$ & 0.05$(\delta/0.1)$ &
0.01+0.07$(\delta/0.1)$ \\ \hline
\end{tabular}
\end{center}
\caption{As in Table~\ref{tb:yexamples1}, but with $\Lambda=770$~MeV.}
\vspace{0.2cm}
\label{tb:yexamples2}
\end{table}
\begin{table}[t]
\begin{center}
\begin{tabular}{|c|c|c|c|c|} \hline
$N$ & $M$ & $Y^8_1$ & $Y^8_2$ &
$Y^{27}$ \\ \hline
3 & 200 & -0.38 & 0.32 & 0.02 \\ \hline
3 & 350 & -0.22 & 0.26 & 0.01 \\ \hline
3 & 500 & -0.27 & 0.27 & 0.05 \\ \hline
2 & 200 & -0.24 & 0.25 & 0.05 \\ \hline
2 & 350 & -0.01 & 0.13 & 0.02 \\ \hline
2 & 500 & -0.14 & 0.13 & -0.03 \\ \hline
0 & 200 & -0.23$(\delta/0.1)$ & 0.17$(\delta/0.1)$ &
-0.05+0.07$(\delta/0.1)$ \\ \hline
0 & 350 & 0.00$(\delta/0.1)$ & 0.05$(\delta/0.1)$ &
0.00+0.07$(\delta/0.1)$ \\ \hline
0 & 500 & 0.14$(\delta/0.1)$ & -0.02$(\delta/0.1)$ &
0.19+0.07$(\delta/0.1)$ \\ \hline
\end{tabular}
\end{center}
\caption{As in Table~\ref{tb:yexamples1}, but with $\Lambda=550$~MeV.}
\vspace{0.2cm}
\label{tb:yexamples3}
\end{table}

\begin{table}[t]
\begin{center}
\begin{tabular}{|c|c|c|c|c|c|c|c|} \hline
$N$ & $M$ & $\cM$ & $\cM(M)$ & $\cM_\infty$ &
$A$ & $A(M)$ & $A_\infty$ \\ \hline
3 & 200 & 433 & 434 & 500 & 0.56 & 0.61 & 1.0 \\ \hline
3 & 350 & 433 & 445 & 500 & 0.56 & 0.73 & 1.0 \\ \hline
3 & 500 & 433 & 500 & 500 & 0.56 & 1.00 & 1.0 \\ \hline
2 & 200 & 499 & 501 & 612 & 0.66 & 0.74 & 1.5 \\ \hline
2 & 350 & 499 & 518 & 612 & 0.66 & 0.95 & 1.5 \\ \hline
2 & 500 & 499 & 612 & 612 & 0.66 & 1.50 & 1.5 \\ \hline
\end{tabular}
\end{center}
\caption{Comparison of values for $\cM$ and $A$, calculated
from Eqs.~(\ref{XLIMIT}) ($\cM$, $A$), (\ref{X}) ($\cM(M)$,
$A(M)$) and (\ref{CALMLIMIT}) ($\cM_\infty$, $A_\infty$), for
$\delta=0.18$, $\alpha=0$ and $M_{SS}=500$~MeV.  $M$ and $\cM$ are
in MeV.}
\vspace{0.2cm}
\label{tb:calm}
\end{table}
\begin{table}[t]
\begin{center}
\begin{tabular}{|c|c|c|c|c|} \hline
$N$ & $M$ & $X^\gamma$ & $Y^\gamma_1$ &
$Y^\gamma_2$ \\ \hline
3 & 200 & -0.08 & 0.09 & 0.14 \\ \hline
3 & 350 & -0.12 & 0.15 & 0.21 \\ \hline
3 & 500 & -0.17 & 0.19 & 0.27 \\ \hline
2 & 200 & -0.10 & 0.11 & 0.17 \\ \hline
2 & 350 & -0.16 & 0.18 & 0.26 \\ \hline
2 & 500 & -0.21 & 0.24 & 0.34 \\ \hline
0 & 200 & -0.11 & 0.14 & 0.16 \\ \hline
0 & 350 & -0.23 & 0.29 & 0.35 \\ \hline
0 & 500 & -0.34 & 0.40 & 0.52 \\ \hline
\end{tabular}
\end{center}
\caption{The one-loop corrections $X^\gamma$ and $Y^\gamma_{1,2}$ for
$2M_K^2/3=M_\pi^2=M^2$, $M_{SS}=500$~MeV, $\Lambda=1000$~MeV,
$\delta=0.18$ for $N=2,3$, $M$ in MeV.  The results do not depend
on $\delta$ for $N=0$.}
\vspace{0.2cm}
\label{tb:examples4}
\end{table}

It turns out that the one-loop correction $X^{27}$ for $[K\to\pi]_{27}$
is very large.  However, just as one defines the kaon $B$ parameter
$B_K$ in the case of $K^0-{\overline{K}}^0$ mixing,
it makes sense to consider the ratio
of this matrix element with its value evaluated by vacuum
saturation or for large $N_c$, which is proportional to $(M^2f^2)_{\rm phys}$.
In one-loop partially quenched ChPT, this can be expressed
in terms of $M$ and $f$ as \cite{glpq}
\ba
(M^2f^2)_{\rm phys}&\!=&\!M^2f^2\left(1+X_{\rm vs}+\frac{2}{(\pi f)^2}
\left(M^2\lambda_8+NM_{SS}^2\lambda_6\right)\right)\;,
\label{MF} \\
X_{\rm vs}&\!=&\!-\frac{1}{(4\pi f)^2}
\left[\frac{2}{3}(\cM^2L(M)-AM^2(2L(M)-1))\right. \nonumber \\
&&\hspace{5.0truecm}\left.
+2N\Mbar^2(L(\Mbar)-1)\right]\;. \nonumber
\ea
For the ratio
\be
B_{27}=\frac{[K\to\pi]_{27}}{(M^2f^2)_{\rm phys}} \label{BTS}
\ee
the relevant one-loop correction is 
\be
\hX^{27}=X^{27}-X_{\rm vs}
=-\frac{1}{(4\pi f)^2}
\left[2M^2(3L(M)-2)\right]\;. \label{XHAT}
\ee
In our examples, we will always consider the quantity $\hX^{27}$ instead
of $X^{27}$.  From Eqs.~(\ref{KPITSPLET}) and (\ref{MF}) we see that
$\hX^{27}$ is independent of $N$, $M_{SS}$ and the $\eta'$ parameters.

In Tables~\ref{tb:xexamples1} to \ref{tb:yexamples3} and Table 
\ref{tb:examples4} we computed the one-loop corrections
for various choices of the valence-meson mass $M$ and three values of the 
$\overline{MS}$ scale $\Lambda=1$ GeV, 770 MeV and 550 MeV.  
For $N=0$, we have set the parameters $\alpha$ and $\gamma^8_{12,}$
equal to zero, for simplicity.  In Figs.
2, 4 and 6, we show the dependence of the three largest octet one-loop 
corrections, $X^8_1$ and $Y^8_{1,2}$ on $M$ and $M_{SS}$,
for $N=2$ and large $M_{\eta'}$.   We do not show similar plots for
$X^8_2$ and $Y^{27}$ because these one-loop corrections are typically
much smaller.  We also do not show a plot for $\hX^{27}$, because it
only depends on $M$ and not on $M_{SS}$, nor on any of the $\eta'$
parameters.

These examples illustrate various points:
\begin{itemize}
\item One-loop corrections can be substantial, and 
will have to be taken into account 
in order to obtain a reliable estimate for $\alpha^8_{1,2}$ 
and $\alpha^{27}$ from the lattice.  It may happen that the $O(p^4)$
LECs have values such that the combined, scale independent contribution
of $O(p^4)$ non-analytic terms and 
counterterms are smaller than the non-analytic
terms alone at a given scale $\Lambda\leq 1$~GeV, 
improving the convergence of ChPT.  One would hope this to be the
case, especially for $\hX^{27}$, which is very large at the
larger values of $M$ and $\Lambda$, and for $Y^8_1$ at larger
$\Lambda$. This issue can be investigated on the lattice. 
\item The size of the one-loop corrections grows with
increasing $\Lambda$.  In particular,
they are relatively small for $\Lambda=m_\eta\approx 550$~MeV.
However, without further knowledge of $O(p^4)$ LECs, it
is unnatural to use $m_\eta$ as the scale, because the $\eta$ is
itself a Goldstone boson, with mass very close to the meson masses
we are considering here.  In the absence of information on $O(p^4)$
LECs, we believe that using higher values for $\Lambda$ gives a
better {\it a priori} estimate of the size of $O(p^4)$ effects.
If, however, the values of $O(p^4)$ LECs turn out to be such that
$\Lambda=550$~MeV gives the better estimate, one-loop corrections
would be reasonably small, and one-loop ChPT should be applicable
in the computation of $\alpha^8_{1,2}$ and $\alpha^{27}$ from
the lattice at realistic values of the quark masses. 
\item The fact that, in a number of cases, a one-loop correction becomes
larger (and, in fact, diverges)
with decreasing $M$ is a consequence of the fact that
we do not vary $M_{SS}$ at the same time, {\it i.e.} of (partial)
quenching.  These ``enhanced" chiral logarithms can be seen very
clearly in the figures in the region $M<M_{SS}$, toward smaller
$M$.  Chiral logarithms are typically smaller near the line $M=M_{SS}$,
for fixed $M+M_{SS}$.
Enhanced chiral logarithms appear in all quantities except
$\hX^{27}$. 
\item  
So far, we have used the large-$M_{\eta'}$ results, {\it i.e.}
those of Subsect. 4.2, for all our examples.
It is interesting to compare the $N\ne 0$ values obtained at a finite
$\eta'$ mass with those obtained for large $M_{\eta'}$. 
In order to do this we chose $\delta=0.18$ for the finite-$M_{\eta'}$
case.  (In the real world $\delta\approx 0.18$;
for quenched QCD $\delta\approx 0.1$ with a large error
\cite{delta}).  We set the other $\eta'$-related
parameters $\alpha$ and $\gamma^8_{1,2}$ equal to zero, simply
because not much is known about their values (but see below for
some remarks on contributions proportional to these
parameters).

For $N=2$, we show the difference between large and finite $M_{\eta'}$
in Figs. 3, 5 and 7, for $X^8_1$ and $Y^8_{1,2}$.  We show the
ratios $[1+X({\rm finite}\ M_{\eta'})]/[1+X({\rm large}\ %
M_{\eta'})]$, with $X= X^8_1,\, Y^8_2$, and the difference
$Y^8_1({\rm finite}\ M_{\eta'})-Y^8_1({\rm large}\ M_{\eta'})$
(there is no tree-level contribution proportional to $\alpha^8_1$). 
The first thing to be 
noticed is that there are still enhanced chiral logarithms in the
small $M$, large $M_{SS}$ region.  This is because the coefficients
of the enhanced chiral logarithms depend on $M_{\eta'}$.  

In the small $M$, $M_{SS}$ region, the ratios are close to one
(or, the difference is close to zero), as
one would expect if both $M$ and $M_{SS}$ are small compared to
$M_{\eta'}$.  For $N=2$, $M_{\eta'}=863$~MeV, and
$M^2_{SS}/M^2_{\eta'}\approx 0.34$ for $M_{SS}=500$~MeV, while for $N=3$,  
$M_{\eta'}=996$~MeV, so that $M^2_{SS}/M^2_{\eta'}\approx 0.25$
for $M_{SS}=500$~MeV.  We point out that for larger meson masses, 
the plots are less meaningful in the region $M>M_{SS}$.
This is a consequence of the
fact that we expanded the finite-$M_{\eta'}$ results in
$M^2/M^2_{\eta'}$,
but not in $M_{SS}^2/M^2_{\eta'}$.  This makes sense if $M<M_{SS}$,
but not for $M\ge M_{SS}$.  Therefore, for valence-quark masses
close to the sea-quark mass, one either has to work consistently
with the large-$M_{\eta'}$ results, or a more general analysis using
Eq.~(\ref{X}) instead of Eq.~(\ref{XLIMIT}) is
needed.  The latter is outside the scope of the present paper,
but for $M^2_{SS}/M^2_{\eta'}\approx 0.25-0.34$ it appears reasonable
to use the large-$M_{\eta'}$ results of Subsect.~4.2. 

As a further illustration of this point, Table \ref{tb:calm} shows
the values of $\cM^2$ and $A$ 
calculated from Eq.~(\ref{XLIMIT}) (3rd, resp. 6th columns),
the ``exact" expression Eq.~(\ref{X}) with degenerate valence
quark masses (4th and 7th columns), and in the limit of
large $M_{\eta'}$, Eq.~(\ref{CALMLIMIT}) (5th and 8th columns).
We see that the variation of values for $\cM^2$ and
$A$ is at most about 20, resp. 50\%. This is not very large from a
practical point of view:  the current best determination of
$m^2_0=\cM^2(N=0)$ in the quenched theory \cite{delta} has an error
of the same order.  We note that for $M=350$~MeV the values
of $\cM$ and $A$ are already much closer to their ``exact"
values $\cM(M)$ and $A(M)$ than $\cM_\infty$ and $A_\infty$.

All this means that, for the value of parameters
chosen in our examples, one could use the simplest possible form
of the chiral logarithms, given in Subsect.~4.2,
to fit numerical results.  Tables \ref{tb:xexamples1} to
\ref{tb:yexamples3} then give the relevant examples of the size
of non-analytic one-loop corrections.
Only with numerical results so
precise that one would be able to determine $\cM^2$ and $A$
with better precision would it be important to take the dependence
on $\eta'$ parameters into account. 
Note again that these conclusions do depend on
the values of the meson masses (and hence quark masses) we
considered in our examples.
\item We also considered the effect of the $\eta'$ couplings
$\gamma^8_{1,2}$.  They contribute to the octet matrix elements at
one loop through the chiral logarithms $X^\gamma$ and $Y^\gamma_{1,2}$.
In Table~\ref{tb:examples4} we show the size of these one-loop corrections
for the choice $\Lambda=1$~GeV.
We see that these quantities are not very small, even though they
vanish in the limit in which the $\eta'$ decouples (for $N\ne 0$).
They are typically smaller than $X^8_{1,2}$ and $Y^8_{1,2}$.
If the couplings $\gamma^8_{1,2}$ themselves are small as well
(compared to $\alpha^8_{1,2}$), it may be possible to ignore these
terms.  But if these couplings are not small, this would mean that
the $\eta'$ does play a role, even if we can use the large-$M_{\eta'}$
partially-quenched expressions of Subsect.~4.2 for the other
chiral logarithms.

The results are not very sensitive to $\alpha$, except in
the quenched case ($N=0$).  We expect that, within the precision
of current lattice computations, the effect of (arbitrarily)
setting $\alpha$, as well as $\gamma^8_{1,2}$, to zero, can be 
accommodated in fits by shifts in the $O(p^4)$ LECs
$C^8_V$ {\it etc.} in Eq.~(\ref{GENERIC}), without
significant impact on the values of the $O(p^2)$
LECs $\alpha^8_{1,2}$ and $\alpha^{27}$.  This can be
checked by fitting lattice results while constraining
these parameters to a few different values.  

\end{itemize}
\newpage
%
%

\begin{figure}
\begin{center}
\hspace*{-2.0cm}
\begin{minipage}[t]{0.12\linewidth}
\vspace*{-4.0cm}
$X_1^8 $
\end{minipage}
\hspace*{-0.5cm}
\leavevmode\epsfxsize=8cm\epsfysize=8cm\epsfbox{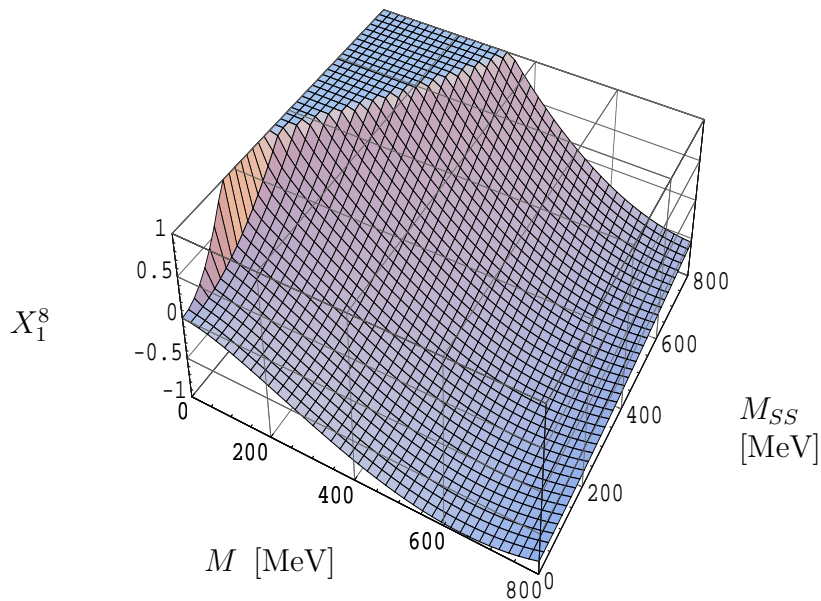}\\[5pt]
\hspace*{2.5cm}
\begin{minipage}[t]{0.12\linewidth}
\vspace*{-1.5cm}
$M\,$ [MeV]
\end{minipage}
\hspace*{5.0cm}
\begin{minipage}[t]{0.12\linewidth}
\vspace*{-3.5cm}
$M_{SS}\,$ [MeV] 
\end{minipage}
\end{center}
\caption{$X_1^8({\rm large}\ M_{\eta'})$ with $\Lambda=1$~GeV,
 as a function of $M$ and $M_{SS}$ in MeV.
\label{figure2}}
\end{figure}

\begin{figure}
\begin{center}
\hspace*{-2.0cm}
\begin{minipage}[t]{0.12\linewidth}
\vspace*{-4.0cm}
$R_1^8 $
\end{minipage}
\hspace*{-0.5cm}
\leavevmode\epsfxsize=8cm\epsfysize=8cm\epsfbox{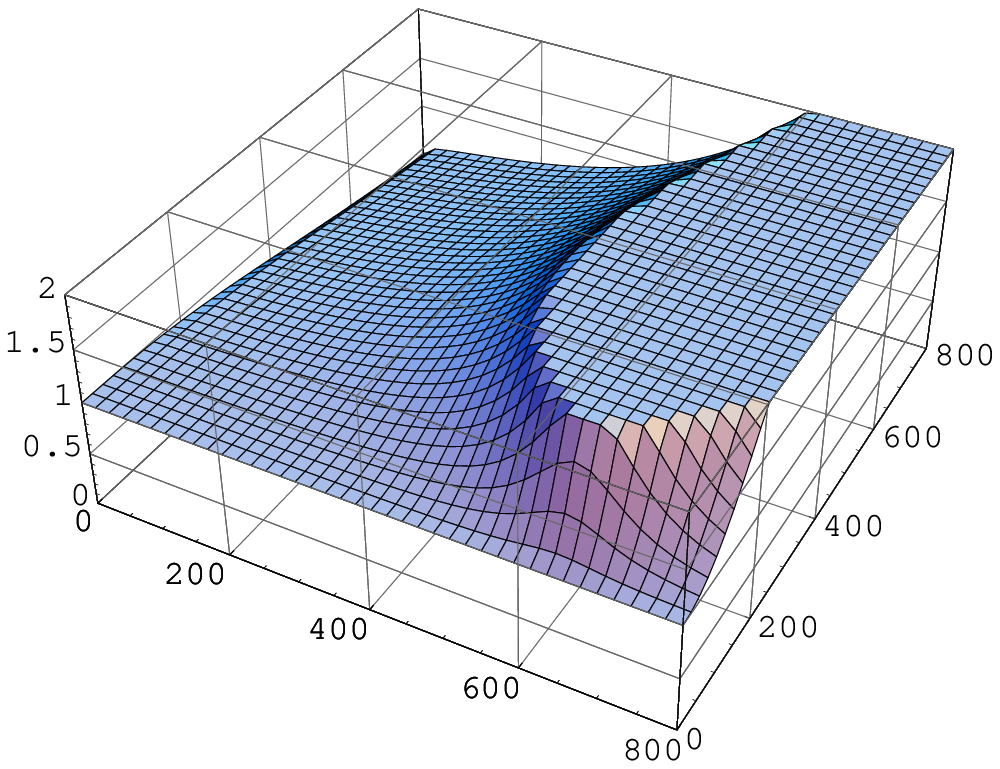}\\[5pt]
\hspace*{2.5cm}
\begin{minipage}[t]{0.12\linewidth}
\vspace*{-1.5cm}
$M\,$ [MeV]
\end{minipage}
\hspace*{5.0cm}
\begin{minipage}[t]{0.12\linewidth}
\vspace*{-3.5cm}
$M_{SS}\,$ [MeV] 
\end{minipage}
\end{center}
\caption{$R^8_1=(1+X_1^8({\rm finite}\ M_{\eta'}))
/(1+X_1^8({\rm large}\ M_{\eta'}))$  with $\Lambda=1$~GeV,
as a function of $M$ and 
$M_{SS}$ in MeV.
\label{figure3}}
\end{figure}
\begin{figure}
\begin{center}
\hspace*{-2.0cm}
\begin{minipage}[t]{0.12\linewidth}
\vspace*{-4.0cm}
$Y_1^8 $
\end{minipage}
\hspace*{-0.5cm}
\leavevmode \epsfxsize=8cm\epsfysize=8cm  \epsfbox{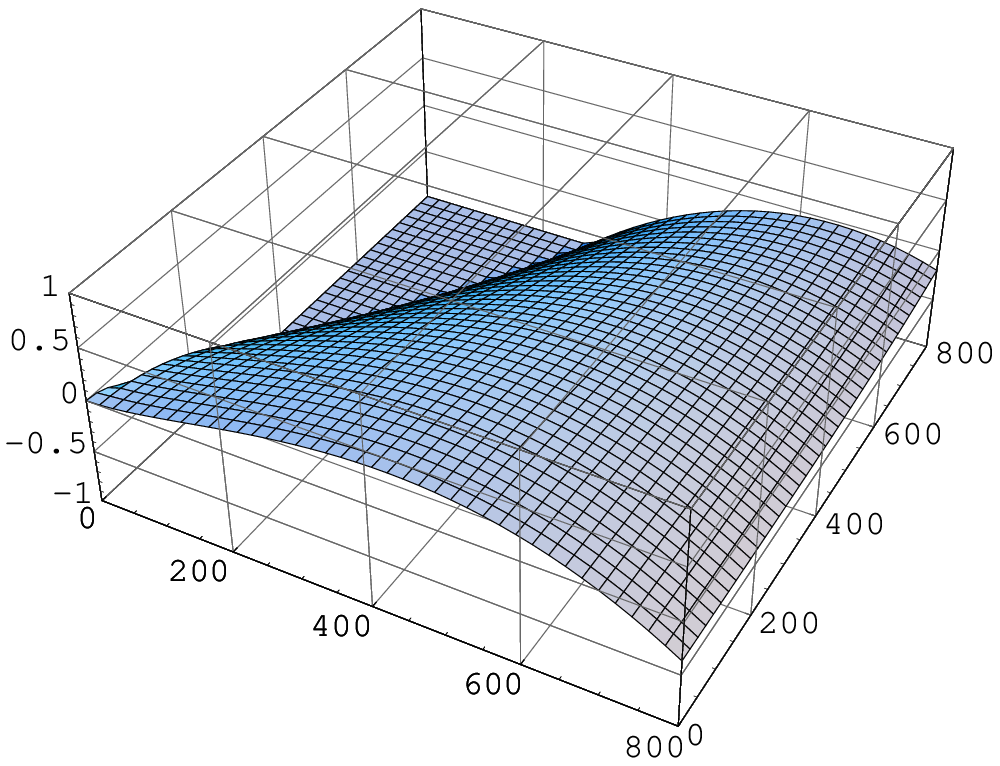}\\[5pt]
\hspace*{2.5cm}
\begin{minipage}[t]{0.12\linewidth}
\vspace*{-1.5cm}
$M\,$ [MeV]
\end{minipage}
\hspace*{5.0cm}
\begin{minipage}[t]{0.12\linewidth}
\vspace*{-3.5cm}
$M_{SS}\,$ [MeV] 
\end{minipage}
\end{center}
\caption{$Y_1^8({\rm large}\ M_{\eta'})$ with $\Lambda=1$~GeV,
as a function of $M$ and $M_{SS}$ in MeV.
\label{figure4}}
\end{figure}
\begin{figure}
\begin{center}
%
%
\leavevmode  \epsfxsize=8cm\epsfysize=8cm  \epsfbox{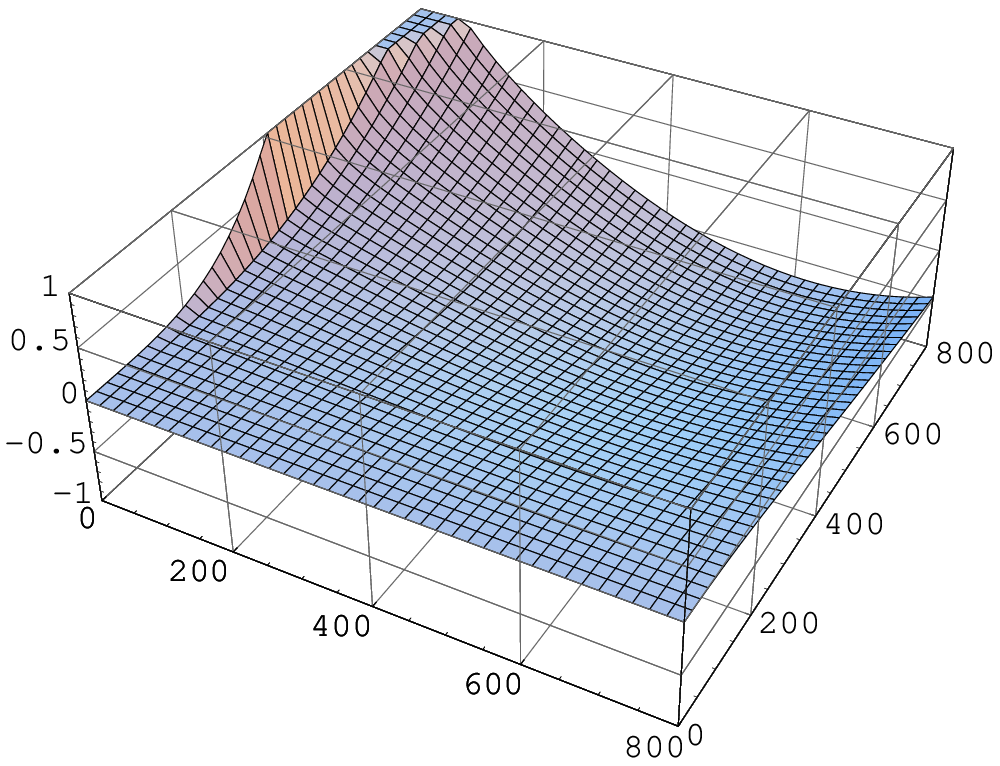}\\[5pt]
\hspace*{2.5cm}
\begin{minipage}[t]{0.12\linewidth}
\vspace*{-1.5cm}
$M\,$ [MeV]
\end{minipage}
\hspace*{5.0cm}
\begin{minipage}[t]{0.12\linewidth}
\vspace*{-3.5cm}
$M_{SS}\,$ [MeV] 
\end{minipage}
\end{center}
\caption{$Y_1^8({\rm finite}\ M_{\eta'})-Y_1^8({\rm large}\ M_{\eta'})$ 
 with $\Lambda=1$~GeV, as a function of $M$ and $M_{SS}$ in MeV.
\label{figure5}}
\end{figure}
\begin{figure}
\begin{center}
\hspace*{-2.0cm}
\begin{minipage}[t]{0.12\linewidth}
\vspace*{-4.0cm}
$Y_2^8 $
\end{minipage}
\hspace*{-0.5cm}
\leavevmode \epsfxsize=8cm\epsfysize=8cm  \epsfbox{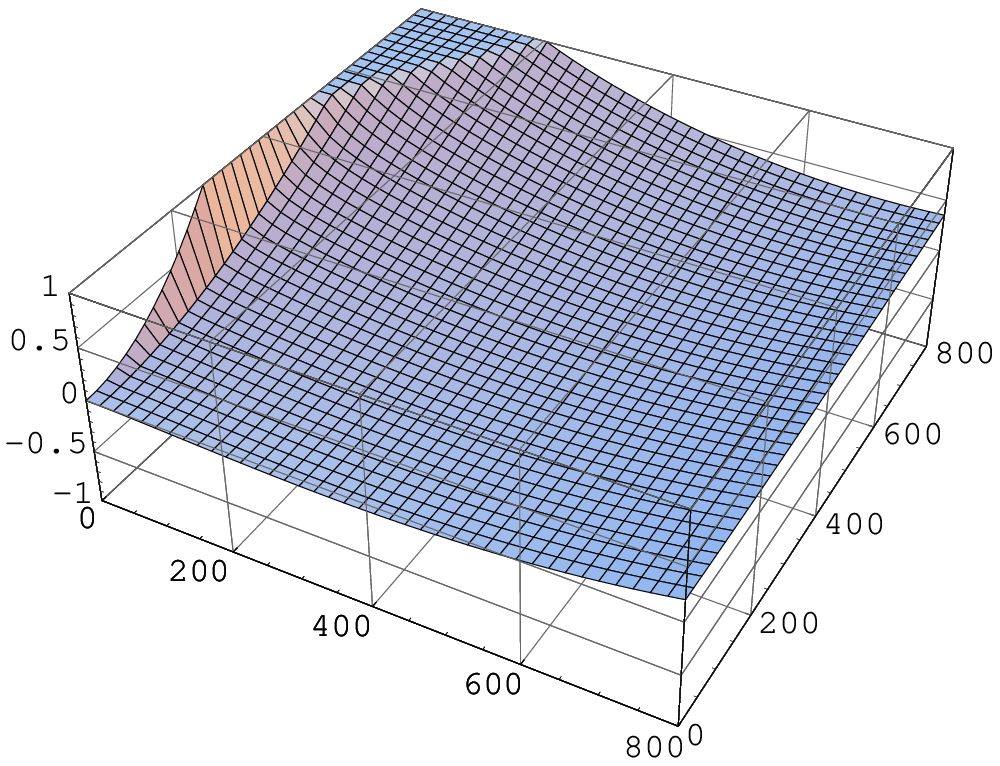}\\[5pt]
\hspace*{2.5cm}
\begin{minipage}[t]{0.12\linewidth}
\vspace*{-1.5cm}
$M\,$ [MeV]
\end{minipage}
\hspace*{5.0cm}
\begin{minipage}[t]{0.12\linewidth}
\vspace*{-3.5cm}
$M_{SS}\,$ [MeV] 
\end{minipage}
\end{center}
\caption{$Y_2^8({\rm large}\ M_{\eta'})$  with $\Lambda=1$~GeV,
as a function of $M$ and $M_{SS}$ in MeV.
\label{figure6}}
\end{figure}
\begin{figure}
\begin{center}
\hspace*{-2.0cm}
\begin{minipage}[t]{0.12\linewidth}
\vspace*{-4.0cm}
$R_2^8 $
\end{minipage}
\hspace*{-0.5cm}
\leavevmode  \epsfxsize=8cm\epsfysize=8cm \epsfbox{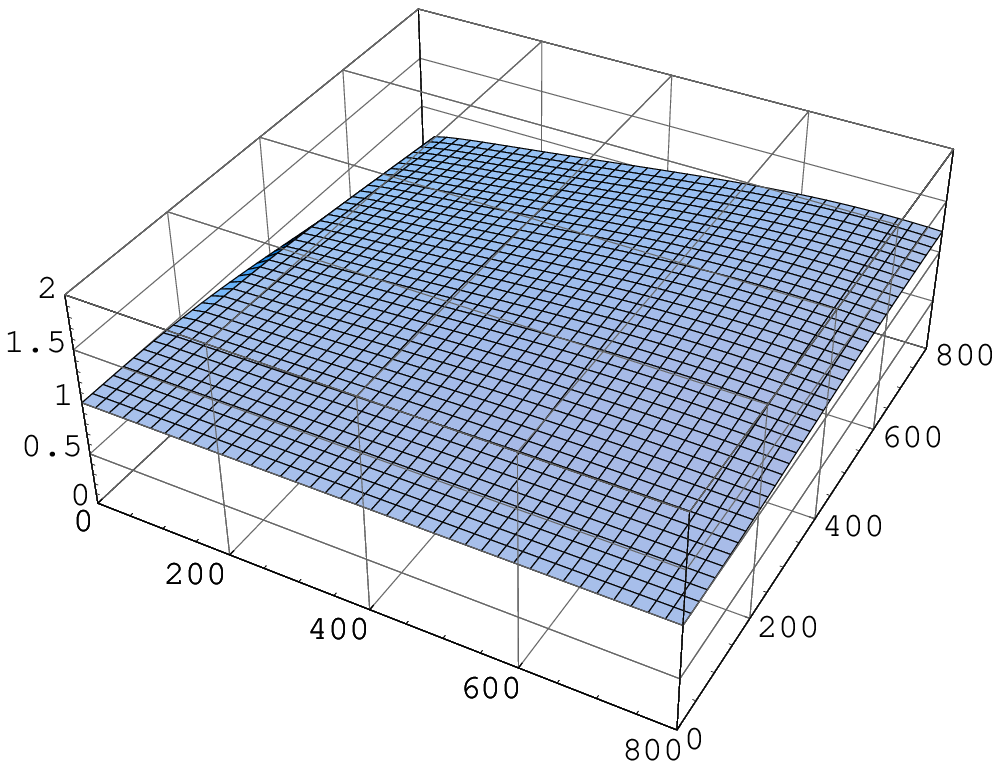}\\[5pt]
\hspace*{2.5cm}
\begin{minipage}[t]{0.12\linewidth}
\vspace*{-1.5cm}
$M\,$ [MeV]
\end{minipage}
\hspace*{5.0cm}
\begin{minipage}[t]{0.12\linewidth}
\vspace*{-3.5cm}
$M_{SS}\,$ [MeV] 
\end{minipage}
\end{center}
\caption{$R^8_2=(1+Y_2^8({\rm finite}\ M_{\eta'}))/(1+Y_2^8
({\rm large}\ M_{\eta'}))$ with $\Lambda=1$~GeV, as a function of $M$ and 
$M_{SS}$ in MeV.
\label{figure7}}
\end{figure}
%
%
%
\section{Conclusion}
\secteq{6}
We presented a complete analysis of $K\to\pi$ and $K\to 0$
weak matrix elements in one-loop ChPT for partially quenched
QCD with $N$ degenerate sea quarks.  For $K\to\pi$ we took 
the valence quarks degenerate in mass, while for $K\to 0$ they
are kept non-degenerate in order to get a non-trivial result.

Three cases have been considered. The first is a partially quenched theory 
with a valence-meson mass much smaller than the $\eta'$ mass, but
with arbitrary value of $M_{SS}/M_{\eta'}$, so that 
an expansion in powers of $M^2_{kk}/M^2_{\eta'}$ is allowed. The second choice
corresponds to the large $ M_{\eta'}$ limit where the $\eta'$ decouples,
so that we can also expand in $M^2_{SS}/M^2_{\eta'}$.
The third case is the quenched theory obtained for $N=0$ or 
$M_{SS}\to\infty$.  For completeness, we also included results
for the unquenched theory.
 
These results should be useful for extracting the values of
the LECs $\alpha^8_{1,2}$ and $\alpha^{27}$ from Lattice QCD.
As we emphasized already in the
Introduction, these are interesting quantities in
their own right. Estimates extracted from experiment
exist, with which lattice results can be compared.
The matrix elements considered here
are the simplest weak matrix elements which can be used
for this goal.  The expressions to be used in fits to lattice
data are given in Eq.~(\ref{GENERIC}), in which $X^8_{1,2}$,
$X^{27}$, $Y^8_{1,2}$, $Y^{27}$, $X^\gamma$ and $Y^\gamma_{1,2}$
represent one-loop corrections (chiral logarithms), and $C^{8,27}_{V,S}$, 
$D^8_{V,S}$ and $D^{27}$ are linear combinations of $O(p^4)$ 
LECs.  They can be read off from the
explicit one-loop results in Sects.~(4.1-4).

{}From our numerical examples in Sect.~5 it is clear that one-loop
expressions will be needed for typical values of quark masses
used in present lattice computations.  Contributions from
$O(p^4)$ operators, represented by the LECs $C^{8,27}_{V,S}$,
$D^8_{V,S}$ and $D^{27}$, will also need to be included.  

Theoretically, values for these $O(p^4)$ LECs can also be
extracted from lattice computations.  However, realistically,
we expect that such estimates would have large uncertainties,
both as a consequence of the typical statistics of present lattice
computations, as well as because of uncertainties
related to the role of the $\eta'$ discussed in more detail in Sect.~5. 
We note that the $O(p^4)$ LEC $\beta^{27}_1$ can be
accessed directly by a computation of the ratio of the
$[K\to 0]_{27}$ and the $K^0-{\overline{K}}^0$-mixing matrix
elements ({\it cf.} Eq.~(\ref{GENERIC})).

In practice, for reasonably small values of the sea-quark mass
(of order less than one-half times the strange-quark mass),
it may be possible to use the partially quenched results 
for large $M_{\eta'}$, given in Subsect.~4.2.  In this case
the $\eta'$ decouples, and therefore all
dependence on $\eta'$ parameters, $m_0$, $\alpha$ and $\gamma^8_{1,2}$
is removed, making the analysis simpler.  In addition, it is
only in this limit that estimates obtained for $O(p^4)$
LECs can be directly compared to those of the real world,
provided that one chooses $N=3$ sea quarks.  For a more
detailed discussion, see Sects.~3 and 5.   The more
general results for the case that the sea-quark mass is larger,
comparable to the $\eta'$ mass, but the valence-quark mass is 
still small enough, are given in Subsect.~4.1.

A completely quenched
lattice computation (for which the relevant results are
in Subsect.~4.3) should be useful for assessing the feasibility
of this approach.  An $N=2$ computation, in combination with
a quenched computation, could give insight into the dependence
of the LECs on the number of light flavors.  However, since
we do not know the functional form of the $N$ dependence of the
(finite part of the) LECs, an $N=3$
computation will be needed to obtain estimates without
an uncontrolled systematic error.

The emphasis of this paper is on the extraction of reliable
numbers for $\alpha^8_1$ and $\alpha^{27}$ from lattice
computations.  While these $O(p^2)$ LECs are interesting,
because of the availability of phenomenological estimates,
the final aim of such Lattice QCD computations would be
to convert these numbers into quantitative estimates
of the $\Delta I=1/2$ and $\Delta I=3/2$ $K\to 2\pi$ decay
amplitudes.  This can be done using ChPT, and complete $O(p^4)$
formulae for doing so are given in Ref.~\cite{bpp}.  Assuming
that $O(p^4)$ ChPT is precise enough, the largest uncertainty
arises because of the poor knowledge of all needed $O(p^4)$ LECs.
Many of these, as we discussed in Sect.~5, cannot even in
principle be determined 
from the $K\to\pi$ and $K\to 0$ matrix elements.  (More
$O(p^4)$ LECs are accessible through the $K^0-\overline{K}^0$
and $K\to 2\pi$ matrix
elements with both pions at rest \cite{mgep}, the
computation of which can also serve as a check on 
lattice results for $\alpha^8_1$ and $\alpha^{27}$.)
One would
have to resort either to the use of available phenomenological
information \cite{kmw1}, or to theoretical estimates based
on arguments such as large $N_c$ or
 models (for recent discussions see Refs.~\cite{bpp,bp}).

In addition, it is well known that final-state interactions are responsible 
for a large enhancement of the $I=0$ $K\to \pi\pi$ amplitude 
\cite{kmw1,pp,MANY}. In that case it could be necessary to resum those 
effects instead of relying on an $O(p^4)$ ChPT calculation. A possible way 
of resumming final-state interactions has recently been proposed in 
Ref.~\cite{pp}.
\bigskip
\subsubsection*{Acknowledgements}
We would like to thank Claude Bernard, Steve Sharpe as well as Akira
Ukawa and other members of the CP-PACS collaboration for 
very useful discussions.  MG would like to thank the Physics
Departments of the Universit\`a ``Tor Vergata," Rome, the
Universitat Autonoma, Barcelona, and the University of Washington,
Seattle, for hospitality.  MG is supported in part by the 
US Department of Energy, and EP by the Ministerio de Educa\c cion
y Cultura of Spain.

\end{document}